\documentclass[aps,prb,times,twocolumn,amsmath,amssymb,superscriptaddress,floatfix]{revtex4}

\usepackage{color}

\usepackage{graphicx}
\usepackage{subfigure}
\usepackage{bbold}
\usepackage{verbatim}
\usepackage{float}
\usepackage{enumerate}
\usepackage{amsfonts}
\usepackage{multirow}
\usepackage[colorlinks,bookmarks=false,citecolor=blue,linkcolor=red,urlcolor=blue]{hyperref}

\begin{document}

%%%%%%%%%%%%%%%%%%%%%%%%%%%%%%%%%%%%%%%%%%%%%%%%%%%%%%

\title{

Topological aspects of symmetry breaking in triangular-lattice Ising antiferromagnets

}

%%%%%%%%%%%%%%%%%%%%%%%%%%%%%%%%%%%%%%%%%%%%%%%%%%%%%%

\author{Andrew Smerald}

\affiliation{Institute of Physics, Ecole Polytechnique F{\'e}d{\'e}rale de Lausanne (EPFL), CH-1015 Lausanne, Switzerland}

\author{Sergey Korshunov\footnote{deceased}}

\affiliation{L.~D.~Landau Institute for Theoretical Physics, Kosygina 2, Moscow 119334, Russia}

\author{Fr{\'e}d{\'e}ric Mila}
\affiliation{Institute of Physics, Ecole Polytechnique F{\'e}d{\'e}rale de Lausanne (EPFL), CH-1015 Lausanne, Switzerland}

%%%%%%%%%%%%%%%%%%%%%%%%%%%%%%%%%%%%%%%%%%%%%%%%%%%%%%

\date{\today}

%%%%%%%%%%%%%%%%%%%%%%%%%%%%%%%%%%%%%%%%%%%%%%%%%%%%%%

\begin{abstract}  
Using a specially designed Monte Carlo algorithm with directed loops, we investigate the
triangular lattice Ising antiferromagnet with coupling beyond nearest neighbour.
We show that the first-order transition from the stripe state to the paramagnet 
can be split, giving rise to an intermediate nematic phase in which algebraic correlations coexist with a broken symmetry.
Furthermore, we demonstrate the emergence of several properties of a more topological nature such as
 fractional edge excitations in the stripe state, the proliferation of double domain walls in the nematic phase,
and the Kasteleyn transition between them.
%Furthermore, we demonstrate the topological nature of several aspects of the problem:
%The phase transitions are best determined using winding numbers rather than order parameters, the low-temperature
%stripe state supports fractional edge excitations, the intermediate phase is characterized by the proliferation
%of double domain walls, and the transition between them is a Kasteleyn one.
Experimental implications are briefly discussed. 
%: a degeneracy between different topological sectors, a gap to all bulk fluctuations and fractional edge excitations.
%In the intermediate nematic state
%
\end{abstract}

%\pacs{
%
%}

\maketitle

The triangular-lattice Ising antiferromagnet (TLIAF) is the archetypal model of frustration.
Ground states of the nearest-neighbour (n.n.) model obey the local constraint that triangles cannot host three equivalent Ising spins, and it follows that there is an extensive entropy\cite{wannier50,houtappel50}.
This results in a critical state, characterised by algebraic correlations between the spins\cite{stephenson64,stephenson70}.

In reality, interactions are rarely limited to n.n., and a more realistic Hamiltonian takes the form,
\begin{align}
\mathcal{H}_{\sf Is} = \sum_{(i,j)} J_{ij} \sigma_i \sigma_j,
\label{eq:HIs}
\end{align}
where $\sigma_i = \pm 1$ and $J_{ij}>0$.
This model is experimentally relevant in a diverse range of systems, including artificial dipolar magnets\cite{mengotti09}, materials such as Ba$_3$CuSb$_2$O$_9$ where electrically charged dumbbells act as Ising degrees of freedom\cite{nakatsuji12,smerald15}, trapped ions\cite{britton12,senko14}, frustrated Coulomb liquids\cite{mahmoudian15}, Josephson junction arrays\cite{korshunov05-prl} and absorbed monolayers\cite{villain83}.

In spite of its ubiquity, this model has received limited attention. %
The difficulty in analysing $\mathcal{H}_{\sf Is}$ [Eq.~\ref{eq:HIs}] arises from the critical nature of the n.n. ground-state manifold, which is very sensitive to perturbation, and in the presence of further-neighbour coupling the model is not amenable to an analytic solution. 
Besides, in the limit $J_1\to \infty$ as compared to the other characteristic energy scales of the problem ($J_2$, $J_3$, etc.), Monte Carlo (MC) simulations based on the Metropolis algorithm are unable to reach the ground state, and the problem of freezing remains even when this constraint is relaxed, for example in the case of dipolar interactions\cite{rossler01}. 
\begin{figure}[t]
\centering
\includegraphics[width=0.49\textwidth]{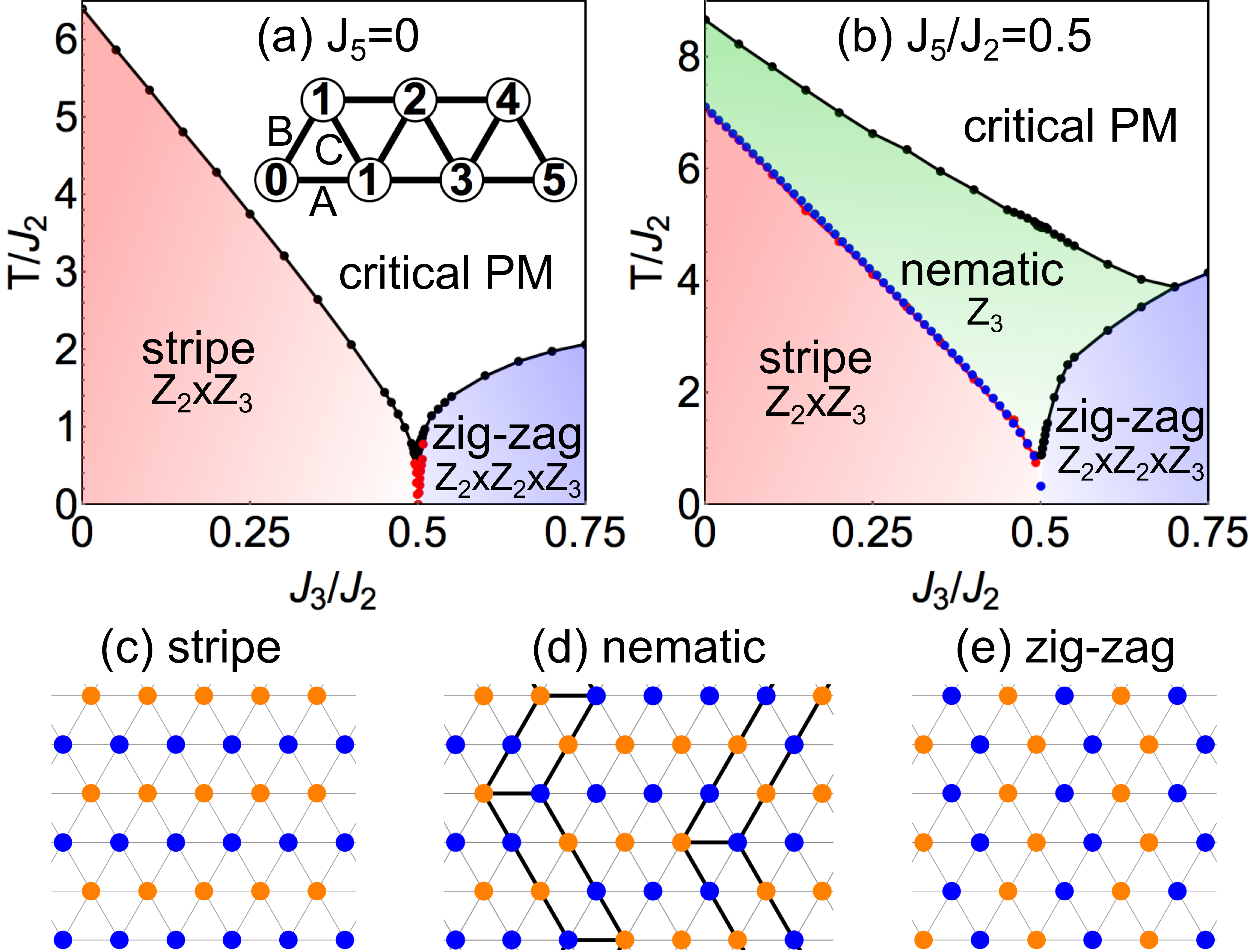}
\caption{\footnotesize{
Representative phase diagrams of the TLIAF with $J_1 \to \infty$, determined by MC simulation.
All phase boundaries were determined from the winding number.
(a) In the presence of $J_2$ and $J_3$ interactions there are 3 possible phases, and all transitions are 1st order.
(Inset) Illustration of 1st to 5th neighbours of the central site and bond labelling.
(b) In the presence of a $J_5$ interaction an intermediate nematic state is revealed.
Blue (red) dots show the position of a 2nd order Kasteleyn transition calculated analytically\cite{korshunov05} (with MC) [see Appendix~\ref{app:Tddw}]. 
(c) Stripe phase.
(d) Nematic phase with fluctuating double domain walls (shown in black) [see Appendix~\ref{app:J1J2J3J5} for more detailed snapshots].
(e) Zig-zag phase.
}}
\label{fig:phasediags}
\end{figure}

The current understanding of the properties of $\mathcal{H}_{\sf Is}$ is based on estimates of the energy and entropy of different types of extended defects.
This results in the prediction that the broken  $Z_2 \times Z_3$ symmetry of the low-temperature stripe state \cite{metcalf74,kaburagi74,korshunov05}
can be restored either in a single first-order transition, or via a pair of transitions, where the low-temperature, $Z_2$-restoring transition is second order and the higher-temperature, $Z_3$-restoring transition is first order\cite{korshunov05}.
When the transition is split, an intermediate phase of nematic type is revealed, and it is characterised by a set of fluctuating double domain walls\cite{korshunov05} (ddw).

In this letter, we show that the difficulty in simulating the TLIAF with MC arises from the topological structure of the n.n. manifold of ground states.
This can be resolved by employing a specially designed worm algorithm [see Appendix~\ref{app:wormalg}] that allows one to travel through the different topological sectors present in the $J_1 \rightarrow  \infty$ limit, and by using the topological winding number rather than the order parameters to map out the phase
diagrams. 
The resulting phase diagrams of two representative models (see Fig.~\ref{fig:phasediags})
are in excellent agreement with the predictions of Ref.~[\onlinecite{korshunov05}], including the stabilization of an intermediate nematic phase 
for large enough fifth-neighbour coupling.
We also show that the hidden topological nature of the model leads to a number of new insights, including fractional edge excitations, and the Kasteleyn nature of the phase transition between the stripe and the nematic phases.\cite{kasteleyn63}

\begin{figure}[t]
\centering
\includegraphics[width=0.45\textwidth]{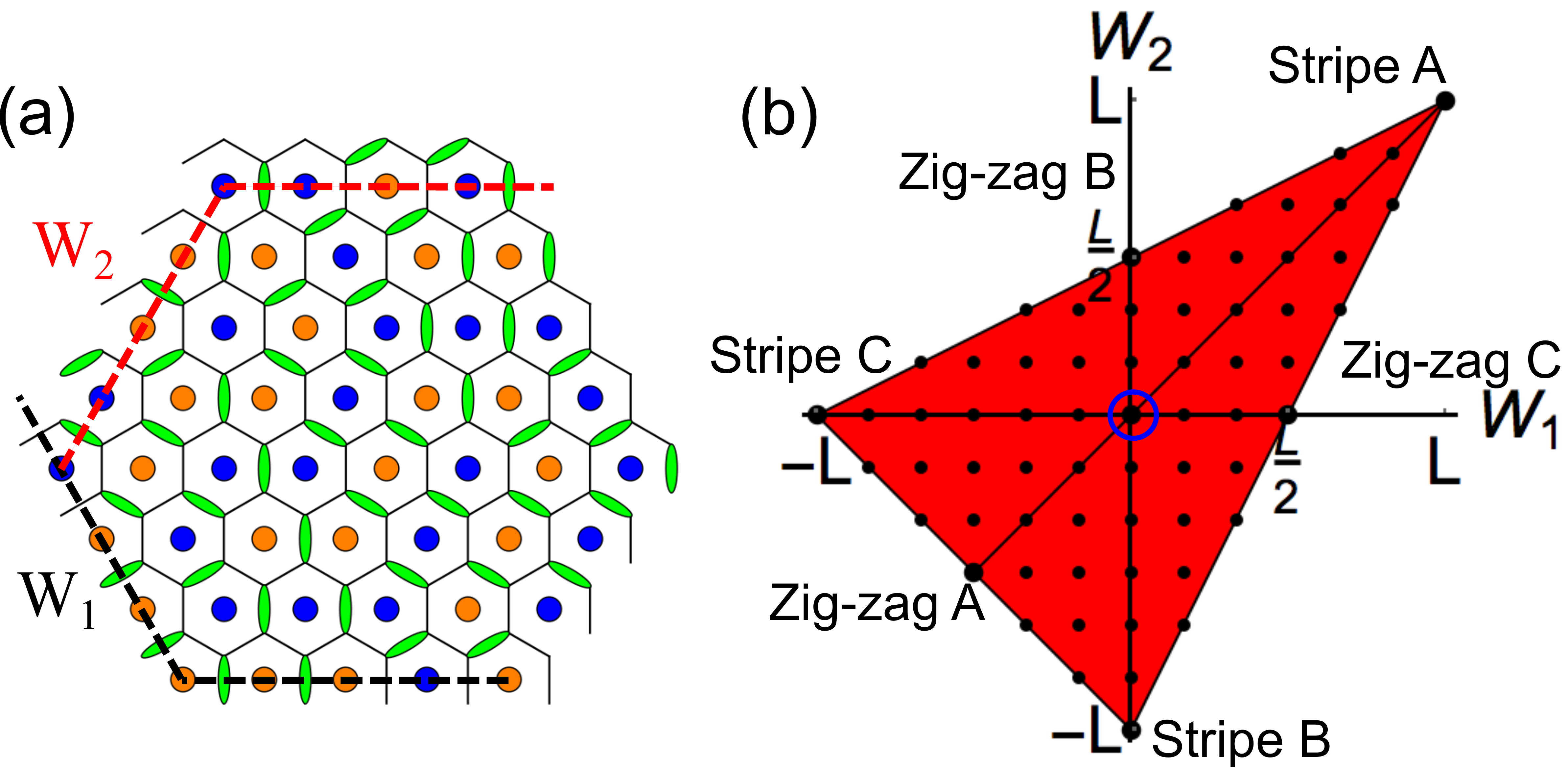}
\caption{\footnotesize{
Winding number sectors of the TLIAF.
(a) An Ising configuration within the nearest-neighbour ground state manifold can be mapped onto a dimer covering of the dual honeycomb lattice.
The dimer crossings of the reference lines are used to calculate the winding number $W=(W_1,W_2)$.
(b) The allowed winding number sectors, illustrated for $L=12$.
The $W=(0,0)$ sector (circled in blue) has a macroscopic degeneracy, while the sectors containing stripe (zig-zag) states have a 2 (4)-fold degeneracy.
}}
\label{fig:windingsecs}
\end{figure}
\begin{figure}[t]
\centering
\includegraphics[width=0.49\textwidth]{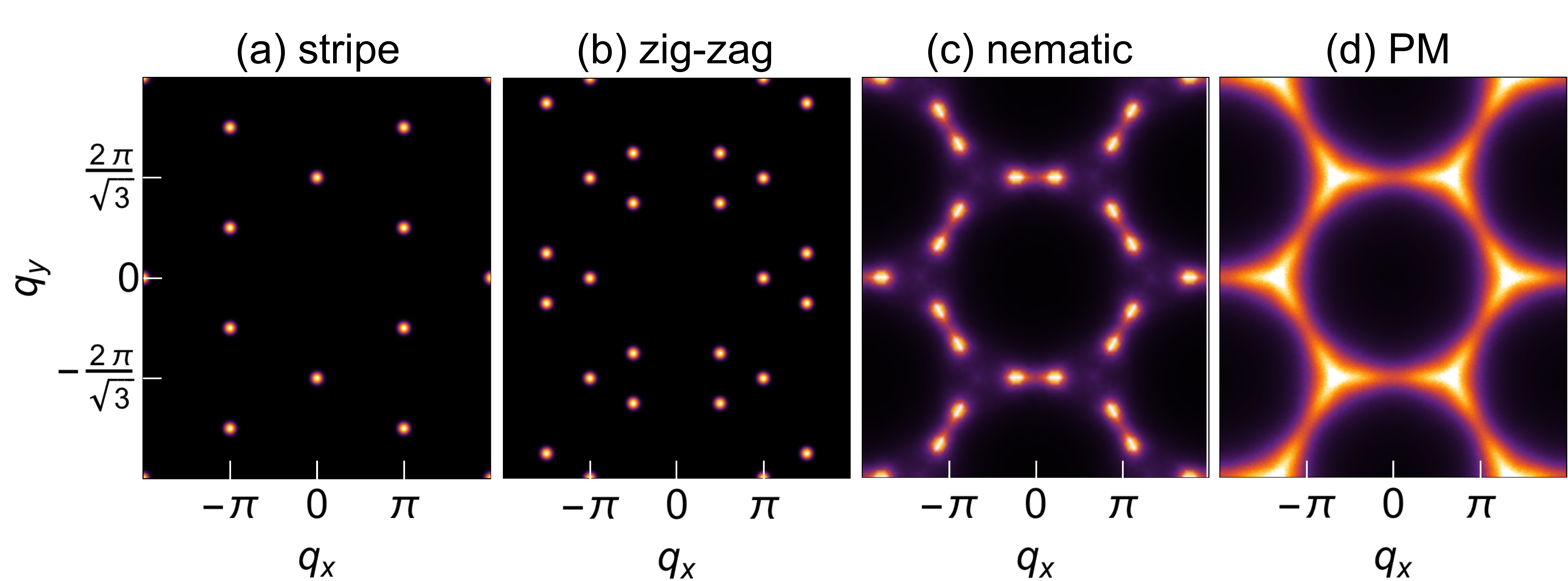}
\caption{\footnotesize{
Representative plots of the structure factor in the 4 phases (see Fig.~\ref{fig:phasediags}).
(a) In the stripe phase there are a set of Bragg peaks, and similarly (b) in the zig-zag phase.
(c) The nematic phase is characterised by pairs of peaks, with power-law singularities, and the peak splitting is proportional to the density of double domain walls.
(d) The critical paramagnet also has peaks with power-law singularities, but at different positions compared to the nematic state.
}}
\label{fig:strucfac}
\end{figure}
%

%%%%%%%%%%%%%%%%%%%%%%%%%%%%%%%%%%%%%%%%%%%%%%%%%%%%%%
% NN model and winding number
%%%%%%%%%%%%%%%%%%%%%%%%%%%%%%%%%%%%%%%%%%%%%%%%%%%%%%

%{\it Topological sectors and algorithm:} 
%
We start by reviewing the topological properties of the n.n. ground-state manifold.
This can be split into topological sectors by defining a pair of winding numbers $W=(W_1,W_2)$.
A useful first step is to map the TLIAF onto a dimer model on the dual honeycomb lattice, in which a dimer is placed on each honeycomb bond that separates two equivalent Ising spins\cite{kasteleyn63} (see Fig.~\ref{fig:windingsecs}).
Two reference lines are defined on the triangular lattice and for each honeycomb bond crossing the reference line the associated winding number is augmented by $\mp 1/3$ in the absence of a dimer and $\pm2/3$ in the presence of a dimer.
The sign is determined by defining a direction in which the reference line should be crossed, splitting the honeycomb lattice into two interpenetrating sublattices {\sf 1} and {\sf 2} and taking the upper sign if the bond direction is {\sf 1}$\to${\sf 2} and the lower sign for {\sf 2}$\to${\sf 1}.
The n.n. manifold is dominated by the $W=(0,0)$ sector, and in the thermodynamic limit the ratio of configurations in this sector compared to the total number of configurations tends to 0.996 [see Appendix~\ref{app:windsec}].

Let us now consider the effect of further-neighbor couplings on this degenerate ground state manifold (or equivalently the limit $J_1 \rightarrow \infty$).
A positive $J_2$ selects 3 stripe states which belong to the topological sector $W=(L,L)$ for stripes parallel to {\sf A} bonds, $W=(0,-L)$ for {\sf B}-stripes and $W=(-L,0)$ for {\sf C}-stripes. 
These topological sectors are as far as possible from the dominant sector $W=(0,0)$, and up to the $Z_2$ degeneracy associated with global spin flips, they contain only these states (see Fig.~\ref{fig:windingsecs}).
The situation is similar for the zig-zag ground states realized for $J_3/J_2>1/2$ [see Appendix~\ref{app:J1J2J3}]. 
Starting from high temperature, these states are thus out of reach of a single spin flip algorithm. So we have developed a sophisticated worm algorithm
with non-local updates that allow the system to change topological sector.
Such updates involve identifying loops of alternating dimer-covered and empty bonds, and exchanging the two, thus flipping all the spins contained within the loop.\cite{zhang09}
In addition, we found that it was necessary in practice to direct the creation of the loops using all further-neighbour interactions, and this results in rejection-free updates [see Appendix~\ref{app:wormalg} for a detailed description of the algorithm].

%%%%%%%%%%%%%%%%%%%%%%%%%%%%%%%%%%%%%%%%%%%%%%%%%%%%%%
% stripe state - topological properties
%%%%%%%%%%%%%%%%%%%%%%%%%%%%%%%%%%%%%%%%%%%%%%%%%%%%%%

%{\it The phase diagrams:}
To map out the phase diagram of the models with $J_1 \rightarrow\infty$, the most efficient way was to keep track of the non-analyticities of the temperature dependent winding number defined by \mbox{$W_{\sf max} = \mathrm{max}(|W_1|,|W_2|,|W_2-W_1|)$}.
Including only $J_2$ and $J_3$ the resulting phase diagram consists of three phases: a high temperature paramagnetic phase, and two low-temperature ordered phases, a stripe phase for $J_3/J_2<1/2$, and a zig-zag phase for $J_3/J_2>1/2$. 
The general phase diagram can potentially include an additional intermediate phase of nematic character, as shown in Fig.~\ref{fig:phasediags} for $J_5/J_2=0.5$.
The nature of the various phases can be revealed by looking at snapshots [see Appendix~\ref{app:J1J2J3J5}]. 
A more precise characterization of direct experimental 
relevance is provided by the structure factor
\mbox{$S({\bf q}) =  \sum_{ij} \left \langle \sigma_i \sigma_j \right \rangle e^{i{\bf q}\cdot {\bf r}}$},
with \mbox{${\bf r} = {\bf r}_i - {\bf r}_j$}. 
Simulations are shown in Fig.~\ref{fig:strucfac}.
The ordered phases have Bragg peaks, 
while the nematic phase has power-law singularities.

\begin{figure}[t]
\centering
\includegraphics[width=0.38\textwidth]{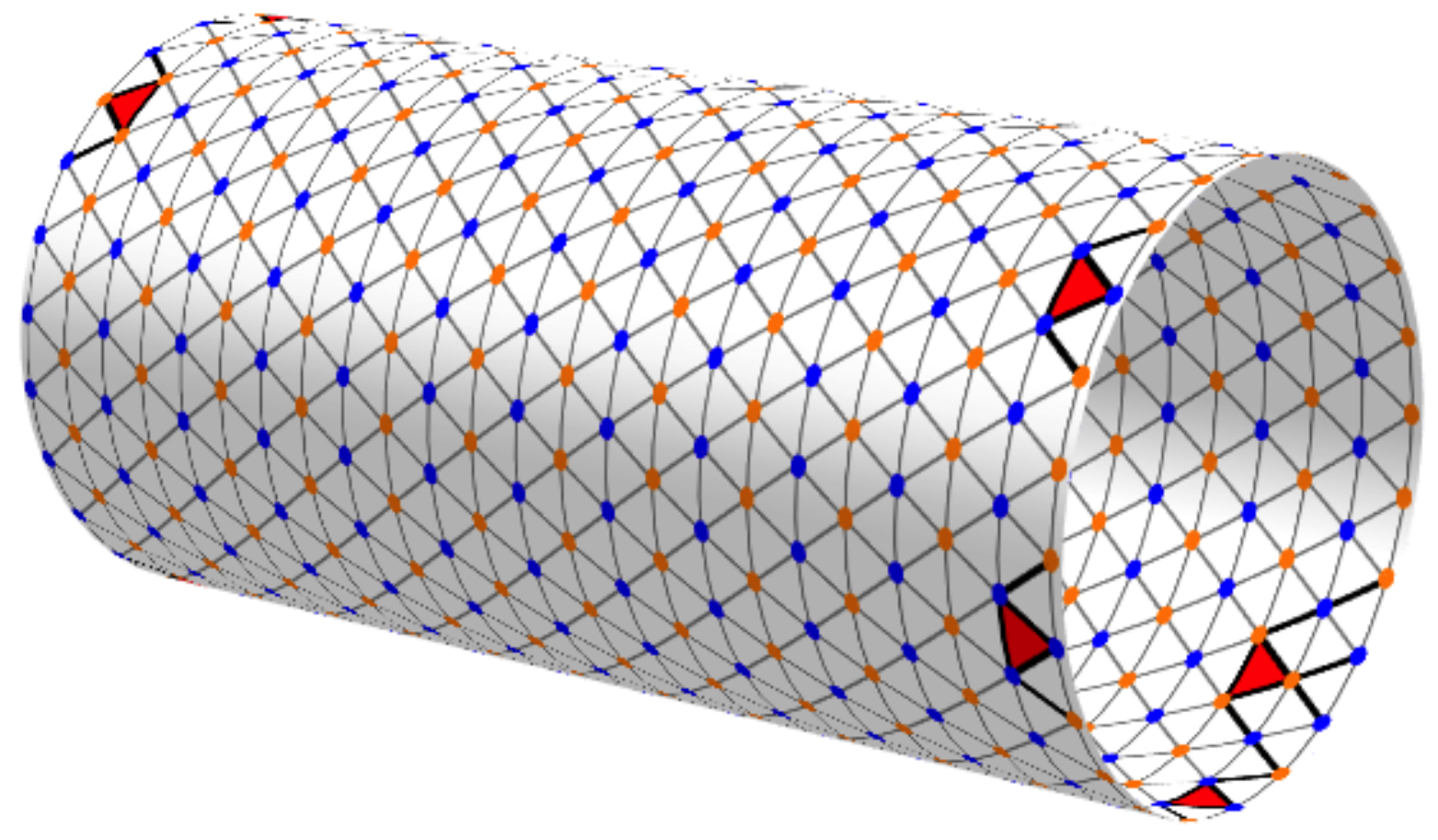}
\caption{\footnotesize{
Fractional edge excitations in the low-temperature stripe state.
While the stripe state is fluctuationless in the bulk, defect triangles (red) can be created at open boundaries at an energy cost of $2J_2-8J_3+4J_4$.
The snapshot shown is taken from a Monte Carlo simulation of the $J_1$-$J_2$ model performed on a cylinder at $T=1.5 J_2$ (for comparison $T_1=6.39J_2$).
}}
\label{fig:edgefluccylinder}
\end{figure}

%{\it The phases:} 
We next discuss in more detail the various phases, starting with the stripe phase.
This simple looking structure has remarkable properties. 
Firstly, the state is fluctuationless at all $T$.
Local excitations involve the creation of pairs of defect triangles, with an energy cost of $4J_1$, while non-local excitations involve the formation of ddw excitations that wind the system \cite{korshunov05}, with an energy cost proportional to $J_2L$. 
Both of these types of excitations are excluded, the local ones due to the condition $J_1 \to \infty$ and the non-local ones by taking the thermodynamic limit $L\to \infty$. 
The zig-zag phase has very similar properties.
%
%The temperature at which the gap to formation of ddws closes, $T_{\sf ddw}$, can be calculated exactly\cite{korshunov05,supmat}, and this is the temperature at which there is a transition to the nematic state.
%
%Alternatively, in the absence of $J_5$, the transition to the critical paramagnet occurs at $T_1 <T_{\sf ddw}$.
%
%This type of gapped, classical low-temperature state is reminiscent of the behaviour of the Kasteleyn model\cite{kasteleyn63,jaubert08}. 
%
However, the stripe state does support fractional edge excitations that are energetically forbidden from  penetrating the bulk.
We have performed simulations with open boundary conditions in order to study these [see Appendix~\ref{app:edges}], and a representative configuration is shown in Fig.~\ref{fig:edgefluccylinder}.
For boundaries orientated parallel to the stripe direction, defect triangles can be created without a $J_1$-energy cost.
In fact they require an energy of only $2J_2-8J_3+4J_4$, and thus there will be a thermally activated population.
Since in the bulk defect triangles are constrained to appear in pairs, these are fractional excitations.
Local dynamics allows these defect triangles to propagate freely along the boundary, but penetration into the bulk is penalized by an energy cost proportional to the penetration depth.

% 

%%%%%%%%%%%%%%%%%%%%%%%%%%%%%%%%%%%%%%%%%%%%%%%%%%%%%%
% intermediate nematic state
%%%%%%%%%%%%%%%%%%%%%%%%%%%%%%%%%%%%%%%%%%%%%%%%%%%%%%

\begin{figure}[t]
\centering
\includegraphics[width=0.4\textwidth]{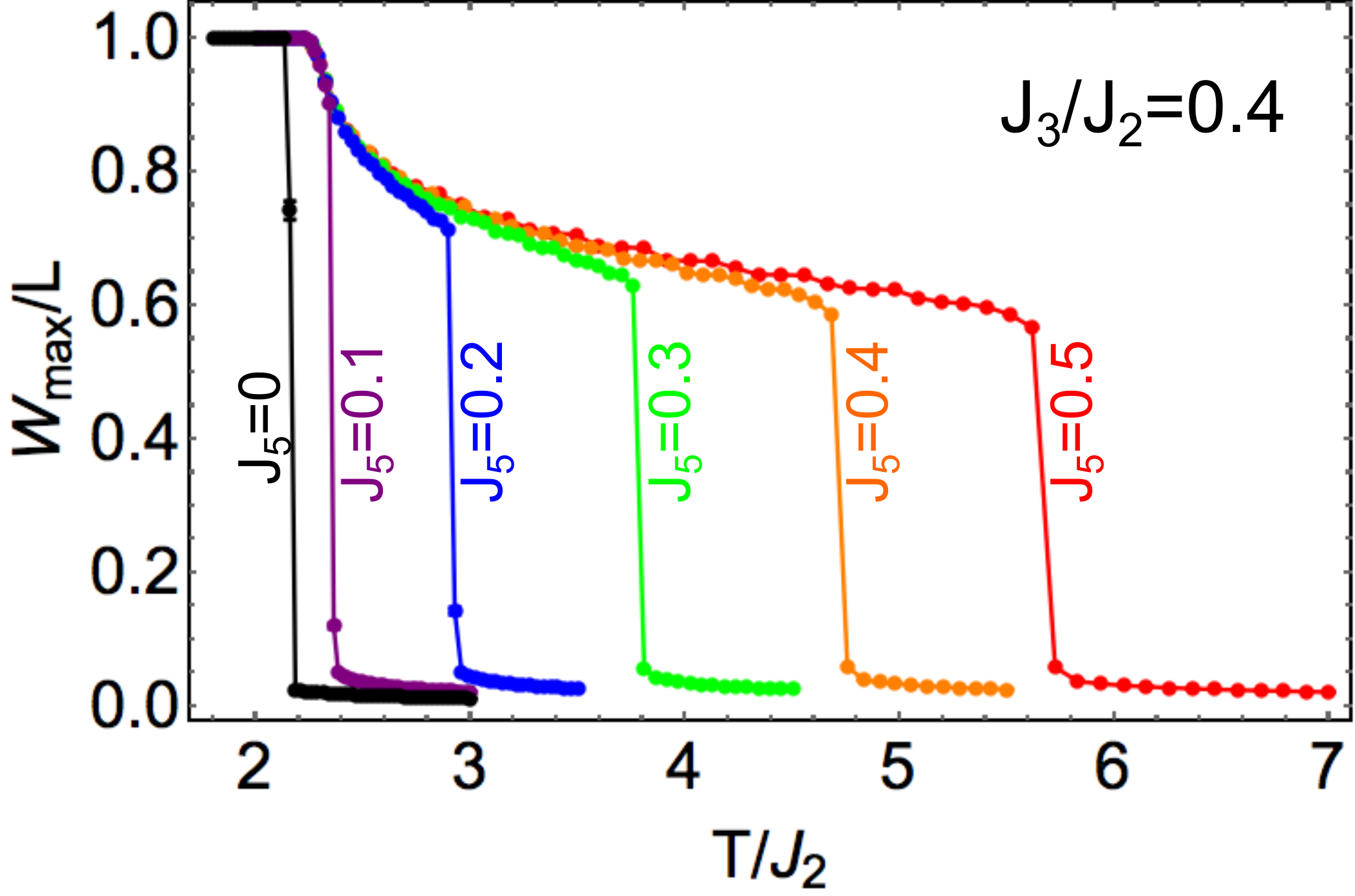}
\caption{\footnotesize{
Temperature dependence of the winding number with $J_2=1$, $J_3=0.4$ and variable $J_5$.
Monte Carlo simulations on an $L=96$ cluster are used to measure \mbox{$W_{\sf max} = \mathrm{max}(|W_1|,|W_2|,|W_2-W_1|)$}, and errors are typically smaller than the point size.
For $J_5=0$ (black) there is a direct 1st order phase transition from the low-temperature stripe state ($W_{\sf max} = L$) to the  high-temperature critical paramagnet ($W_{\sf max} \to 0$).
For all other values of $J_5$ there is a 2nd order phase transition at $T_{\sf ddw}=2.29$ between the stripe state and a nematic state (variable $W_{\sf max}$), followed by a 1st order transition to the critical paramagnet at higher $T$.
}}
\label{fig:J5winding}
\end{figure}

The intermediate nematic state involves the proliferation of ddw defects (see Fig.~\ref{fig:phasediags}).
The ddws run perpendicular to the direction of the stripes and therefore the nematic state breaks the 6-fold rotational symmetry of the triangular lattice down to a 2-fold symmetry.
% (i.e. a $Z_3$ symmetry is broken).
%
The state is best characterised by the density of ddw, $\nu(T)$, and this is related to the winding number according to \mbox{$\nu(T) = 4(L-W_{\sf max})/3L$}.
%, where \mbox{$W_{\sf max} = \mathrm{max}(|W_1|,|W_2|,|W_2-W_1|)$}.
%
It can be seen in Fig.~\ref{fig:J5winding} that $W_{\sf max}$ varies continuously with temperature.
The ddws do not form a periodic arrangement but instead fluctuate, and the state is critical.

The critical nature of the nematic state can be demonstrated by studying the peaks of the structure factor, $S({\bf q})$.
We find that \mbox{$S({\bf q}_{\sf peak} + \delta {\bf q)} \propto |\delta {\bf q}|^{\tau-2}$}, where $\tau$ is a temperature-dependent critical exponent.
This results in algebraic correlations according to \mbox{$S({\bf r}) = \left \langle \sigma_i \sigma_j \right \rangle \propto r^{-\tau}$}, and in the direction perpendicular to the walls one can write \mbox{$S({\bf r}) \propto \cos(\pi r/L) (L/r)^\tau$}.
Simulations show that $0<\tau<1/2$, and it is demonstrated below that $\tau=1/2$ close to $T_{\sf ddw}$.
%In summary the nematic state both breaks a symmetry and displays topologically non-trivial liquid-like properties.
%

%The behaviour of the nematic state can be contrasted with that of the spin-ice phase on the pyrochlore lattice\cite{bramwell01,castelnovo08,castelnovo12}.
%%
%Spin-ice also exhibits algebraic decay of spin correlations and supports a set of fractional excitations, the magnetic monopoles.
%%
%However, the states are clearly very different, since spin-ice is a classical Coulomb liquid described by a gauge theory, while the nematic state in the TLIAF is a liquid crystal with a broken orientational symmetry and a height-model description.
%%
%The spin-ice phase has more in common with the high-temperature critical paramagnet of the TLIAF, and it has been suggested that perturbations of the nearest-neighbour spin-ice model could lead to a variety of unusual, broken-symmetry states\cite{powell11}.

%%%%%%%%%%%%%%%%%%%%%%%%%%%%%%%%%%%%%%%%%%%%%%%%%%%%%%
% phase transition out of the stripe phase
%%%%%%%%%%%%%%%%%%%%%%%%%%%%%%%%%%%%%%%%%%%%%%%%%%%%%%

%
%

%{\it The phase transitions:}

Next we turn to the nature of the phase transitions.
For $J_5=0$ the transition out of the stripe state is first order, and involves an abrupt change of winding number sector.
This can be seen in Fig.~\ref{fig:J5winding}, where in the stripe state $W_{\sf max}=L$, while in the critical paramagnet $W_{\sf max} \to 0$.
For $J_5 \neq 0$ the high-temperature transition from the nematic state to the critical paramagnet is also first order, and involves a transition from an intermediate winding number sector to $W_{\sf max} \to 0$.

More interesting is the low-temperature second-order transition that occurs in the presence of a $J_5$ interaction (see Fig.~\ref{fig:J5winding}).
At this phase transition the Ising $Z_2$ symmetry is broken, and thus one would naively expect that the transition is in the Ising universality class.
However, if one constructs an order parameter based on this symmetry it exhibits a discontinuous jump at the transition.
Furthermore, if one makes a mapping to a dimer model on the honeycomb lattice, the transition remains unchanged, despite the fact that the $Z_2$ symmetry has been discarded.
In fact it is the change of topology and not of symmetry that drives the transition and it is within the Pokrovsky-Talapov universality class.\cite{pokrovsky79,pokrovsky80}

%, and is an example of a commensurate-incommensurate transition\cite{bak82}.

The best way to characterise this transition is as a non-analyticity in the winding number, $W_{\sf max}$, and therefore a divergence of the associated susceptibility.
This occurs at $T_{\sf ddw}$, which can be calculated exactly and is the temperature at which the free-energy of a ddw excitation vanishes.\cite{korshunov05} [see Appendix~\ref{app:Tddw}].
The second-order nature of the transition is ensured by the non-crossing constraint of the ddws, which leads to an entropically driven repulsion.\cite{pokrovsky79,pokrovsky80}

%%%%%%%%%%%%%%%%%%%%%%%%%%%%%%%%%%%%%%%%%%%%%%%%%%%%%%
% free fermion approach to phase transition and beta=1/2
%%%%%%%%%%%%%%%%%%%%%%%%%%%%%%%%%%%%%%%%%%%%%%%%%%%%%%

In order to achieve a quantitative understanding of the critical behaviour, it is useful to make a mapping to a 1d quantum model of fermions\cite{pokrovsky79,pokrovsky80,villain81} [see Appendix~\ref{app:clas-quan}].
A ddw can be identified with a fermion whose imaginary time propagation is parallel to the wall, and the non-crossing condition is encoded in the fermionic anticommutation relation.
The fermionic chemical potential, $\mu$, is related to the temperature of the classical model by $\mu \propto T-T_{\sf ddw}$.
In the fermionic language the second-order phase transition between the nematic and stripe states is a metal-insulator transition.
Close to this transition the density of fermions will be very low, and therefore one can ignore fermion-fermion interactions, which become progressively more important at higher temperatures.
In consequence one finds \mbox{$\nu(T) \propto (T-T_{\sf ddw})^\beta$}, with $\beta=1/2$. 
This critical exponent is typical of the Pokrovsky-Talapov universality class\cite{pokrovsky79,pokrovsky80}, and is clearly not typical of an Ising transition.
A similar analysis of the structure factor leads to the prediction $\tau=1/2$ close to the transition\cite{villain81}.
We have confirmed these predictions by MC simulation [see Appendix~\ref{app:clas-quan}].

If the $J_1 \to \infty$ constraint is relaxed, then for $T>0$ there will be a low but finite density of defect triangles in the bulk of the stripe state.
These have to be created in pairs, and are confined since they are joined by a pair of ddws, and thus the free energy cost grows linearly with their separation [see Appendix~\ref{app:fracexc}].
In contrast, defect triangles are deconfined in the nematic state 
since the free energy of ddws goes to zero on entering the state. In fact, if one of the defect triangles winds around the system, and then the pair annihilates, a pair of ddw has been created. 
We expect that the relaxation of the $J_1 \to \infty$ constraint will cause this transition to cross over to the Ising universality class, but this will only be physically significant in an exponentially suppressed temperature window.\cite{bohr82,schulz82,rutkevich97}
Thus, unless $J_1$ is small, the behaviour 
remains dominated by the Kasteleyn character of the $J_1\rightarrow \infty$ limit. 
%

%{\it Comparison with other models:} 
It is useful to compare and contrast the present findings with those in related systems.
First, a number of states with a similar coexistence of order and liquidity, as in the nematic state, have recently been found in other frustrated magnets\cite{chern11,albuquerque12,borzi13,borzi14,brooks14,powell15,jaubert15}.
There are also close parallels with the floating phases found in systems of gases adsorbed onto a substrate\cite{bak82}.
The transition between the stripe and nematic phases has a lot in common with the Kasteleyn model of dimers on the honeycomb lattice\cite{kasteleyn63} and with spin-ice in a [100] magnetic field, which displays a 3d Kasteleyn transition\cite{fennell05,jaubert08,powell08}.
In particular the second-order transition that we find in the presence of a $J_5$ interaction is within the same Pokrovsky-Talapov universality class as Kasteleyn's model.
However there are a number of important differences.
The model we study is isotropic and therefore the rotational lattice symmetry is broken spontaneously, rather than in the Hamiltonian. 
This isotropy leads to a topological degeneracy in the ground state and the possibility of an intermediate nematic phase.
Also, the $Z_2$ symmetry associated with the Ising spins is not present in the Kasteleyn model, and this leads to the existence of fractional edge states, as well as defect triangles in the bulk when the $J_1 \to \infty$ constraint is relaxed.
The closest analogue to the physics we present here is probably spin-ice with a uniaxial distortion and a 4-spin interaction\cite{powell15}.
In this case the 4-spin interaction splits a single first-order transition into two second-order transitions.
However, the uniaxial distortion breaks most of the symmetry of the pyrochlore lattice by hand, and the physical interpretation of strings is different from the ddws of the TLIAF.

%%%%%%%%%%%%%%%%%%%%%%%%%%%%%%%%%%%%%%%%%%%%%%%%%%%%%%
%height model
%%%%%%%%%%%%%%%%%%%%%%%%%%%%%%%%%%%%%%%%%%%%%%%%%%%%%%

%{\it Splitting condition:}
Finally, let us discuss in what circumstances the transition can be expected to be split.
In Fig.~\ref{fig:phasediags} we have ignored $J_4$ interactions, since this acts against the $J_5$-induced splitting of the transition by reducing the temperature of the first-order line\cite{korshunov05}.
Similar reasoning can be extended to further-neighbour couplings and for a set of interactions that decrease smoothly with distance we expect that a single first-order transition is typical.
We have checked that this is the case for long-range dipolar interactions.
However, the magnetic exchange interaction does not necessarily lead to couplings that decay smoothly with separation, and can therefore result in systems where the transition is split. 

To show that this is a generic possibility, we have derived a general condition by mapping the Ising degrees of freedom onto height configurations of the [111] face of a crystal with a simple cubic lattice\cite{blote82}.
A continuum free energy can thus be written as\cite{korshunov05} [see Appendix~\ref{app:height}],
\begin{align}
\mathcal{F}[h] &= \int d^2 r \left\{ 
\frac{K_2}{2} (\nabla h)^2  
+ K_3 \prod_{\alpha}  ({\bf e}_\alpha \cdot \nabla) h  \right. \nonumber \\
&\left. + \frac{K_4}{4} (\nabla h)^4 
-V_0 \cos \left[ \frac{2\pi}{3}(h_{\sf str} - h) \right]
\right\},
\label{eq:Fheight}
\end{align}
where $K_2$ is temperature dependent, $K_3$ is related to the energy cost of double domain walls, $K_4$ ensures that the free energy is bounded from below and the last term is a locking potential that favours integer values of the height field.
Here $h_{\sf str}({\bf r})$ is the height configuration in one of the stripe states and ${\bf e}_\alpha$ is a set of 3 unit vectors forming 120$^\circ$ angles with one another.
Analysis of this model shows that the transition is split when [see Appendix~\ref{app:height}],
\begin{align}
\frac{K_3^2}{8K_4} > \frac{144 V_0}{\pi^4 [|\nabla h_{\sf str}| - K_3/(2K_4)]^2}.
\end{align}

%%%%%%%%%%%%%%%%%%%%%%%%%%%%%%%%%%%%%%%%%%%%%%%%%%%%%%
%conclusion 
%%%%%%%%%%%%%%%%%%%%%%%%%%%%%%%%%%%%%%%%%%%%%%%%%%%%%%

In conclusion, we have shown that the physics of the extended TLIAF with large nearest-neighbour coupling, a model of 
direct relevance in several contexts, is remarkably rich, with a phase diagram that can only be properly understood by
invoking both broken symmetry and topological concepts.
Indeed, while all phases can be characterized by their broken symmetry,
several of their properties are more topological in nature, such as the fluctuationless character of the low-temperature
stripe state and its fractional edge excitations, the proliferation of double-domain walls in the intermediate nematic
phase that appears when the first-order transition is split by, for example, a  fifth-neighbour interaction, and the Kasteleyn
transition that separate these phases. Far from being of pure academic interest, the topological aspects of the problem might
actually be the key to understanding the properties of systems in which the development of true symmetry breaking is hampered 
by a purely local dynamics, or by the finite size of the sample. For instance, finite clusters are expected to develop
domain walls to minimize their edge energy if they can reach their ground states, or to support edge excitations if they
cannot. We hope that the present paper will motivate experimental studies along these lines.

%%%%%%%%%%%%%%%%%%%%%%%%%%%%%%%%%%%%%%%%%%%%%%%%%%%%%%

{\it Note.} 
We sadly regret the untimely death of one of our co-authors, Sergey Korshunov, shortly before this manusript was completed.
We sincerely hope that the final presentation meets his very high standards.

{\it Acknowledgments.}   
We are grateful to Ludovic Jaubert for a useful discussion.
We also thank the Swiss National Science Foundation and its SINERGIA network ``Mott physics beyond the Heisenberg model'' for financial support.

\bibliographystyle{apsrev4-1}
\bibliography{bibfile}

\begin{thebibliography}{46}%
\makeatletter
\providecommand \@ifxundefined [1]{%
 \@ifx{#1\undefined}
}%
\providecommand \@ifnum [1]{%
 \ifnum #1\expandafter \@firstoftwo
 \else \expandafter \@secondoftwo
 \fi
}%
\providecommand \@ifx [1]{%
 \ifx #1\expandafter \@firstoftwo
 \else \expandafter \@secondoftwo
 \fi
}%
\providecommand \natexlab [1]{#1}%
\providecommand \enquote  [1]{``#1''}%
\providecommand \bibnamefont  [1]{#1}%
\providecommand \bibfnamefont [1]{#1}%
\providecommand \citenamefont [1]{#1}%
\providecommand \href@noop [0]{\@secondoftwo}%
\providecommand \href [0]{\begingroup \@sanitize@url \@href}%
\providecommand \@href[1]{\@@startlink{#1}\@@href}%
\providecommand \@@href[1]{\endgroup#1\@@endlink}%
\providecommand \@sanitize@url [0]{\catcode `\\12\catcode `\$12\catcode
  `\&12\catcode `\#12\catcode `\^12\catcode `\_12\catcode `\%12\relax}%
\providecommand \@@startlink[1]{}%
\providecommand \@@endlink[0]{}%
\providecommand \url  [0]{\begingroup\@sanitize@url \@url }%
\providecommand \@url [1]{\endgroup\@href {#1}{\urlprefix }}%
\providecommand \urlprefix  [0]{URL }%
\providecommand \Eprint [0]{\href }%
\providecommand \doibase [0]{http://dx.doi.org/}%
\providecommand \selectlanguage [0]{\@gobble}%
\providecommand \bibinfo  [0]{\@secondoftwo}%
\providecommand \bibfield  [0]{\@secondoftwo}%
\providecommand \translation [1]{[#1]}%
\providecommand \BibitemOpen [0]{}%
\providecommand \bibitemStop [0]{}%
\providecommand \bibitemNoStop [0]{.\EOS\space}%
\providecommand \EOS [0]{\spacefactor3000\relax}%
\providecommand \BibitemShut  [1]{\csname bibitem#1\endcsname}%
\let\auto@bib@innerbib\@empty
%</preamble>
\bibitem [{\citenamefont {Wannier}(1950)}]{wannier50}%
  \BibitemOpen
  \bibfield  {author} {\bibinfo {author} {\bibfnamefont {G.~H.}\ \bibnamefont
  {Wannier}},\ }\href {\doibase 10.1103/PhysRev.79.357} {\bibfield  {journal}
  {\bibinfo  {journal} {Phys. Rev.}\ }\textbf {\bibinfo {volume} {79}},\
  \bibinfo {pages} {357} (\bibinfo {year} {1950})}\BibitemShut {NoStop}%
\bibitem [{\citenamefont {Houtappel}(1950)}]{houtappel50}%
  \BibitemOpen
  \bibfield  {author} {\bibinfo {author} {\bibfnamefont {R.}~\bibnamefont
  {Houtappel}},\ }\href {\doibase
  http://dx.doi.org/10.1016/0031-8914(50)90130-3} {\bibfield  {journal}
  {\bibinfo  {journal} {Physica}\ }\textbf {\bibinfo {volume} {16}},\ \bibinfo
  {pages} {425 } (\bibinfo {year} {1950})}\BibitemShut {NoStop}%
\bibitem [{\citenamefont {Stephenson}(1964)}]{stephenson64}%
  \BibitemOpen
  \bibfield  {author} {\bibinfo {author} {\bibfnamefont {J.}~\bibnamefont
  {Stephenson}},\ }\href {\doibase http://dx.doi.org/10.1063/1.1704202}
  {\bibfield  {journal} {\bibinfo  {journal} {Journal of Mathematical Physics}\
  }\textbf {\bibinfo {volume} {5}},\ \bibinfo {pages} {1009} (\bibinfo {year}
  {1964})}\BibitemShut {NoStop}%
\bibitem [{\citenamefont {Stephenson}(1970)}]{stephenson70}%
  \BibitemOpen
  \bibfield  {author} {\bibinfo {author} {\bibfnamefont {J.}~\bibnamefont
  {Stephenson}},\ }\href {\doibase http://dx.doi.org/10.1063/1.1665154}
  {\bibfield  {journal} {\bibinfo  {journal} {Journal of Mathematical Physics}\
  }\textbf {\bibinfo {volume} {11}},\ \bibinfo {pages} {413} (\bibinfo {year}
  {1970})}\BibitemShut {NoStop}%
\bibitem [{\citenamefont {Mengotti}\ \emph {et~al.}(2009)\citenamefont
  {Mengotti}, \citenamefont {Heyderman}, \citenamefont {Bisig}, \citenamefont
  {Fraile~Rodr{\'\i}guez}, \citenamefont {Le~Guyader}, \citenamefont
  {Nolting},\ and\ \citenamefont {Braun}}]{mengotti09}%
  \BibitemOpen
  \bibfield  {author} {\bibinfo {author} {\bibfnamefont {E.}~\bibnamefont
  {Mengotti}}, \bibinfo {author} {\bibfnamefont {L.~J.}\ \bibnamefont
  {Heyderman}}, \bibinfo {author} {\bibfnamefont {A.}~\bibnamefont {Bisig}},
  \bibinfo {author} {\bibfnamefont {A.}~\bibnamefont {Fraile~Rodr{\'\i}guez}},
  \bibinfo {author} {\bibfnamefont {L.}~\bibnamefont {Le~Guyader}}, \bibinfo
  {author} {\bibfnamefont {F.}~\bibnamefont {Nolting}}, \ and\ \bibinfo
  {author} {\bibfnamefont {H.~B.}\ \bibnamefont {Braun}},\ }\href
  {http://scitation.aip.org/content/aip/journal/jap/105/11/10.1063/1.3133202}
  {\bibfield  {journal} {\bibinfo  {journal} {Journal of Applied Physics}\
  }\textbf {\bibinfo {volume} {105}},\ \bibinfo {eid} {113113} (\bibinfo {year}
  {2009})}\BibitemShut {NoStop}%
\bibitem [{\citenamefont {Nakatsuji}\ \emph {et~al.}(2012)\citenamefont
  {Nakatsuji}, \citenamefont {Kuga}, \citenamefont {Kimura}, \citenamefont
  {Satake}, \citenamefont {Katayama}, \citenamefont {Nishibori}, \citenamefont
  {Sawa}, \citenamefont {Ishii}, \citenamefont {Hagiwara}, \citenamefont
  {Bridges}, \citenamefont {Ito}, \citenamefont {Higemoto}, \citenamefont
  {Karaki}, \citenamefont {Halim}, \citenamefont {Nugroho}, \citenamefont
  {Rodriguez-Rivera}, \citenamefont {Green},\ and\ \citenamefont
  {Broholm}}]{nakatsuji12}%
  \BibitemOpen
  \bibfield  {author} {\bibinfo {author} {\bibfnamefont {S.}~\bibnamefont
  {Nakatsuji}}, \bibinfo {author} {\bibfnamefont {K.}~\bibnamefont {Kuga}},
  \bibinfo {author} {\bibfnamefont {K.}~\bibnamefont {Kimura}}, \bibinfo
  {author} {\bibfnamefont {R.}~\bibnamefont {Satake}}, \bibinfo {author}
  {\bibfnamefont {N.}~\bibnamefont {Katayama}}, \bibinfo {author}
  {\bibfnamefont {E.}~\bibnamefont {Nishibori}}, \bibinfo {author}
  {\bibfnamefont {H.}~\bibnamefont {Sawa}}, \bibinfo {author} {\bibfnamefont
  {R.}~\bibnamefont {Ishii}}, \bibinfo {author} {\bibfnamefont
  {M.}~\bibnamefont {Hagiwara}}, \bibinfo {author} {\bibfnamefont
  {F.}~\bibnamefont {Bridges}}, \bibinfo {author} {\bibfnamefont {T.~U.}\
  \bibnamefont {Ito}}, \bibinfo {author} {\bibfnamefont {W.}~\bibnamefont
  {Higemoto}}, \bibinfo {author} {\bibfnamefont {Y.}~\bibnamefont {Karaki}},
  \bibinfo {author} {\bibfnamefont {M.}~\bibnamefont {Halim}}, \bibinfo
  {author} {\bibfnamefont {A.~A.}\ \bibnamefont {Nugroho}}, \bibinfo {author}
  {\bibfnamefont {J.~A.}\ \bibnamefont {Rodriguez-Rivera}}, \bibinfo {author}
  {\bibfnamefont {M.~A.}\ \bibnamefont {Green}}, \ and\ \bibinfo {author}
  {\bibfnamefont {C.}~\bibnamefont {Broholm}},\ }\href {\doibase
  10.1126/science.1212154} {\bibfield  {journal} {\bibinfo  {journal}
  {Science}\ }\textbf {\bibinfo {volume} {336}},\ \bibinfo {pages} {559}
  (\bibinfo {year} {2012})}\BibitemShut {NoStop}%
\bibitem [{\citenamefont {Smerald}\ and\ \citenamefont
  {Mila}(2015)}]{smerald15}%
  \BibitemOpen
  \bibfield  {author} {\bibinfo {author} {\bibfnamefont {A.}~\bibnamefont
  {Smerald}}\ and\ \bibinfo {author} {\bibfnamefont {F.}~\bibnamefont {Mila}},\
  }\href {\doibase 10.1103/PhysRevLett.115.147202} {\bibfield  {journal}
  {\bibinfo  {journal} {Phys. Rev. Lett.}\ }\textbf {\bibinfo {volume} {115}},\
  \bibinfo {pages} {147202} (\bibinfo {year} {2015})}\BibitemShut {NoStop}%
\bibitem [{\citenamefont {Britton}\ \emph {et~al.}(2012)\citenamefont
  {Britton}, \citenamefont {Sawyer}, \citenamefont {Keith}, \citenamefont
  {Wang}, \citenamefont {Freericks}, \citenamefont {Uys}, \citenamefont
  {Biercuk},\ and\ \citenamefont {Bollinger}}]{britton12}%
  \BibitemOpen
  \bibfield  {author} {\bibinfo {author} {\bibfnamefont {J.~W.}\ \bibnamefont
  {Britton}}, \bibinfo {author} {\bibfnamefont {B.~C.}\ \bibnamefont {Sawyer}},
  \bibinfo {author} {\bibfnamefont {A.~C.}\ \bibnamefont {Keith}}, \bibinfo
  {author} {\bibfnamefont {C.~C.~J.}\ \bibnamefont {Wang}}, \bibinfo {author}
  {\bibfnamefont {J.~K.}\ \bibnamefont {Freericks}}, \bibinfo {author}
  {\bibfnamefont {H.}~\bibnamefont {Uys}}, \bibinfo {author} {\bibfnamefont
  {M.~J.}\ \bibnamefont {Biercuk}}, \ and\ \bibinfo {author} {\bibfnamefont
  {J.~J.}\ \bibnamefont {Bollinger}},\ }\href
  {http://dx.doi.org/10.1038/nature10981} {\bibfield  {journal} {\bibinfo
  {journal} {Nature}\ }\textbf {\bibinfo {volume} {484}},\ \bibinfo {pages}
  {489} (\bibinfo {year} {2012})}\BibitemShut {NoStop}%
\bibitem [{\citenamefont {Senko}\ \emph {et~al.}(2014)\citenamefont {Senko},
  \citenamefont {Smith}, \citenamefont {Richerme}, \citenamefont {Lee},
  \citenamefont {Campbell},\ and\ \citenamefont {Monroe}}]{senko14}%
  \BibitemOpen
  \bibfield  {author} {\bibinfo {author} {\bibfnamefont {C.}~\bibnamefont
  {Senko}}, \bibinfo {author} {\bibfnamefont {J.}~\bibnamefont {Smith}},
  \bibinfo {author} {\bibfnamefont {P.}~\bibnamefont {Richerme}}, \bibinfo
  {author} {\bibfnamefont {A.}~\bibnamefont {Lee}}, \bibinfo {author}
  {\bibfnamefont {W.~C.}\ \bibnamefont {Campbell}}, \ and\ \bibinfo {author}
  {\bibfnamefont {C.}~\bibnamefont {Monroe}},\ }\href {\doibase
  10.1126/science.1251422} {\bibfield  {journal} {\bibinfo  {journal}
  {Science}\ }\textbf {\bibinfo {volume} {345}},\ \bibinfo {pages} {430}
  (\bibinfo {year} {2014})}\BibitemShut {NoStop}%
\bibitem [{\citenamefont {Mahmoudian}\ \emph {et~al.}(2015)\citenamefont
  {Mahmoudian}, \citenamefont {Rademaker}, \citenamefont {Ralko}, \citenamefont
  {Fratini},\ and\ \citenamefont {Dobrosavljevi\ifmmode~\acute{c}\else
  \'{c}\fi{}}}]{mahmoudian15}%
  \BibitemOpen
  \bibfield  {author} {\bibinfo {author} {\bibfnamefont {S.}~\bibnamefont
  {Mahmoudian}}, \bibinfo {author} {\bibfnamefont {L.}~\bibnamefont
  {Rademaker}}, \bibinfo {author} {\bibfnamefont {A.}~\bibnamefont {Ralko}},
  \bibinfo {author} {\bibfnamefont {S.}~\bibnamefont {Fratini}}, \ and\
  \bibinfo {author} {\bibfnamefont {V.}~\bibnamefont
  {Dobrosavljevi\ifmmode~\acute{c}\else \'{c}\fi{}}},\ }\href {\doibase
  10.1103/PhysRevLett.115.025701} {\bibfield  {journal} {\bibinfo  {journal}
  {Phys. Rev. Lett.}\ }\textbf {\bibinfo {volume} {115}},\ \bibinfo {pages}
  {025701} (\bibinfo {year} {2015})}\BibitemShut {NoStop}%
\bibitem [{\citenamefont {Korshunov}(2005{\natexlab{a}})}]{korshunov05-prl}%
  \BibitemOpen
  \bibfield  {author} {\bibinfo {author} {\bibfnamefont {S.~E.}\ \bibnamefont
  {Korshunov}},\ }\href {\doibase 10.1103/PhysRevLett.94.087001} {\bibfield
  {journal} {\bibinfo  {journal} {Phys. Rev. Lett.}\ }\textbf {\bibinfo
  {volume} {94}},\ \bibinfo {pages} {087001} (\bibinfo {year}
  {2005}{\natexlab{a}})}\BibitemShut {NoStop}%
\bibitem [{\citenamefont {Villain}\ and\ \citenamefont
  {Gordon}(1983)}]{villain83}%
  \BibitemOpen
  \bibfield  {author} {\bibinfo {author} {\bibfnamefont {J.}~\bibnamefont
  {Villain}}\ and\ \bibinfo {author} {\bibfnamefont {M.}~\bibnamefont
  {Gordon}},\ }\href {\doibase http://dx.doi.org/10.1016/0039-6028(83)90443-0}
  {\bibfield  {journal} {\bibinfo  {journal} {Surface Science}\ }\textbf
  {\bibinfo {volume} {125}},\ \bibinfo {pages} {1 } (\bibinfo {year}
  {1983})}\BibitemShut {NoStop}%
\bibitem [{\citenamefont {R{\"o}{\ss}ler}(2001)}]{rossler01}%
  \BibitemOpen
  \bibfield  {author} {\bibinfo {author} {\bibfnamefont {U.~K.}\ \bibnamefont
  {R{\"o}{\ss}ler}},\ }\href {\doibase http://dx.doi.org/10.1063/1.1358336}
  {\bibfield  {journal} {\bibinfo  {journal} {Journal of Applied Physics}\
  }\textbf {\bibinfo {volume} {89}},\ \bibinfo {pages} {7033} (\bibinfo {year}
  {2001})}\BibitemShut {NoStop}%
\bibitem [{\citenamefont {Korshunov}(2005{\natexlab{b}})}]{korshunov05}%
  \BibitemOpen
  \bibfield  {author} {\bibinfo {author} {\bibfnamefont {S.~E.}\ \bibnamefont
  {Korshunov}},\ }\href {\doibase 10.1103/PhysRevB.72.144417} {\bibfield
  {journal} {\bibinfo  {journal} {Phys. Rev. B}\ }\textbf {\bibinfo {volume}
  {72}},\ \bibinfo {pages} {144417} (\bibinfo {year}
  {2005}{\natexlab{b}})}\BibitemShut {NoStop}%
\bibitem [{sup()}]{supmat}%
  \BibitemOpen
  \href@noop {} {}\bibinfo {note} {See Supplemental Material at {\ldots} for
  further information concerning winding number sectors, the Monte Carlo worm
  algorithm, the phase diagram, the transition temperature $T_{\sf ddw}$, the
  second-order phase transition between the stripe and nematic states, the role
  of defect triangles and the analysis of a height model to understand the
  splitting of the transition. This includes
  Ref.~[\onlinecite{wannier50,syljuasen02,sandvik06,alet06,zhang09,korshunov05,glosli83,takagi95,rastelli05,pokrovsky79,pokrovsky80,villain81,yokoi86,jiang06,jaubert08,powell08,powell09,blote82,bak82,schulz82}]}\BibitemShut
  {NoStop}%
\bibitem [{\citenamefont {Metcalf}(1974)}]{metcalf74}%
  \BibitemOpen
  \bibfield  {author} {\bibinfo {author} {\bibfnamefont {B.}~\bibnamefont
  {Metcalf}},\ }\href {\doibase http://dx.doi.org/10.1016/0375-9601(74)90247-3}
  {\bibfield  {journal} {\bibinfo  {journal} {Physics Letters A}\ }\textbf
  {\bibinfo {volume} {46}},\ \bibinfo {pages} {325 } (\bibinfo {year}
  {1974})}\BibitemShut {NoStop}%
\bibitem [{\citenamefont {Kaburagi}\ and\ \citenamefont
  {Kanamori}(1974)}]{kaburagi74}%
  \BibitemOpen
  \bibfield  {author} {\bibinfo {author} {\bibfnamefont {M.}~\bibnamefont
  {Kaburagi}}\ and\ \bibinfo {author} {\bibfnamefont {J.}~\bibnamefont
  {Kanamori}},\ }\href {http://stacks.iop.org/1347-4065/13/i=S2/a=145}
  {\bibfield  {journal} {\bibinfo  {journal} {Japanese Journal of Applied
  Physics}\ }\textbf {\bibinfo {volume} {13}},\ \bibinfo {pages} {145}
  (\bibinfo {year} {1974})}\BibitemShut {NoStop}%
\bibitem [{\citenamefont {Kasteleyn}(1963)}]{kasteleyn63}%
  \BibitemOpen
  \bibfield  {author} {\bibinfo {author} {\bibfnamefont {P.~W.}\ \bibnamefont
  {Kasteleyn}},\ }\href {\doibase http://dx.doi.org/10.1063/1.1703953}
  {\bibfield  {journal} {\bibinfo  {journal} {Journal of Mathematical Physics}\
  }\textbf {\bibinfo {volume} {4}},\ \bibinfo {pages} {287} (\bibinfo {year}
  {1963})}\BibitemShut {NoStop}%
\bibitem [{\citenamefont {Zhang}\ \emph {et~al.}(2009)\citenamefont {Zhang},
  \citenamefont {Garoni},\ and\ \citenamefont {Deng}}]{zhang09}%
  \BibitemOpen
  \bibfield  {author} {\bibinfo {author} {\bibfnamefont {W.}~\bibnamefont
  {Zhang}}, \bibinfo {author} {\bibfnamefont {T.~M.}\ \bibnamefont {Garoni}}, \
  and\ \bibinfo {author} {\bibfnamefont {Y.}~\bibnamefont {Deng}},\ }\href
  {\doibase http://dx.doi.org/10.1016/j.nuclphysb.2009.01.007} {\bibfield
  {journal} {\bibinfo  {journal} {Nuclear Physics B}\ }\textbf {\bibinfo
  {volume} {814}},\ \bibinfo {pages} {461 } (\bibinfo {year}
  {2009})}\BibitemShut {NoStop}%
\bibitem [{\citenamefont {Pokrovsky}\ and\ \citenamefont
  {Talapov}(1979)}]{pokrovsky79}%
  \BibitemOpen
  \bibfield  {author} {\bibinfo {author} {\bibfnamefont {V.~L.}\ \bibnamefont
  {Pokrovsky}}\ and\ \bibinfo {author} {\bibfnamefont {A.~L.}\ \bibnamefont
  {Talapov}},\ }\href {\doibase 10.1103/PhysRevLett.42.65} {\bibfield
  {journal} {\bibinfo  {journal} {Phys. Rev. Lett.}\ }\textbf {\bibinfo
  {volume} {42}},\ \bibinfo {pages} {65} (\bibinfo {year} {1979})}\BibitemShut
  {NoStop}%
\bibitem [{\citenamefont {Pokrovsky}\ and\ \citenamefont
  {Talapov}(1980)}]{pokrovsky80}%
  \BibitemOpen
  \bibfield  {author} {\bibinfo {author} {\bibfnamefont {V.~L.}\ \bibnamefont
  {Pokrovsky}}\ and\ \bibinfo {author} {\bibfnamefont {A.~L.}\ \bibnamefont
  {Talapov}},\ }\href@noop {} {\bibfield  {journal} {\bibinfo  {journal} {Zh.
  Eksp. Teor. Fiz}\ }\textbf {\bibinfo {volume} {78}},\ \bibinfo {pages} {269}
  (\bibinfo {year} {1980})}\BibitemShut {NoStop}%
\bibitem [{\citenamefont {{Villain, J.}}\ and\ \citenamefont {{Bak,
  P.}}(1981)}]{villain81}%
  \BibitemOpen
  \bibfield  {author} {\bibinfo {author} {\bibnamefont {{Villain, J.}}}\ and\
  \bibinfo {author} {\bibnamefont {{Bak, P.}}},\ }\href {\doibase
  10.1051/jphys:01981004205065700} {\bibfield  {journal} {\bibinfo  {journal}
  {J. Phys. France}\ }\textbf {\bibinfo {volume} {42}},\ \bibinfo {pages} {657}
  (\bibinfo {year} {1981})}\BibitemShut {NoStop}%
\bibitem [{\citenamefont {Bohr}(1982)}]{bohr82}%
  \BibitemOpen
  \bibfield  {author} {\bibinfo {author} {\bibfnamefont {T.}~\bibnamefont
  {Bohr}},\ }\href {\doibase 10.1103/PhysRevB.25.6981} {\bibfield  {journal}
  {\bibinfo  {journal} {Phys. Rev. B}\ }\textbf {\bibinfo {volume} {25}},\
  \bibinfo {pages} {6981} (\bibinfo {year} {1982})}\BibitemShut {NoStop}%
\bibitem [{\citenamefont {Schulz}\ \emph {et~al.}(1982)\citenamefont {Schulz},
  \citenamefont {Halperin},\ and\ \citenamefont {Henley}}]{schulz82}%
  \BibitemOpen
  \bibfield  {author} {\bibinfo {author} {\bibfnamefont {H.~J.}\ \bibnamefont
  {Schulz}}, \bibinfo {author} {\bibfnamefont {B.~I.}\ \bibnamefont
  {Halperin}}, \ and\ \bibinfo {author} {\bibfnamefont {C.~L.}\ \bibnamefont
  {Henley}},\ }\href {\doibase 10.1103/PhysRevB.26.3797} {\bibfield  {journal}
  {\bibinfo  {journal} {Phys. Rev. B}\ }\textbf {\bibinfo {volume} {26}},\
  \bibinfo {pages} {3797} (\bibinfo {year} {1982})}\BibitemShut {NoStop}%
\bibitem [{\citenamefont {Rutkevich}(1997)}]{rutkevich97}%
  \BibitemOpen
  \bibfield  {author} {\bibinfo {author} {\bibfnamefont {S.~B.}\ \bibnamefont
  {Rutkevich}},\ }\href {http://stacks.iop.org/0305-4470/30/i=11/a=017}
  {\bibfield  {journal} {\bibinfo  {journal} {Journal of Physics A:
  Mathematical and General}\ }\textbf {\bibinfo {volume} {30}},\ \bibinfo
  {pages} {3883} (\bibinfo {year} {1997})}\BibitemShut {NoStop}%
\bibitem [{\citenamefont {Chern}\ \emph {et~al.}(2011)\citenamefont {Chern},
  \citenamefont {Mellado},\ and\ \citenamefont {Tchernyshyov}}]{chern11}%
  \BibitemOpen
  \bibfield  {author} {\bibinfo {author} {\bibfnamefont {G.-W.}\ \bibnamefont
  {Chern}}, \bibinfo {author} {\bibfnamefont {P.}~\bibnamefont {Mellado}}, \
  and\ \bibinfo {author} {\bibfnamefont {O.}~\bibnamefont {Tchernyshyov}},\
  }\href {\doibase 10.1103/PhysRevLett.106.207202} {\bibfield  {journal}
  {\bibinfo  {journal} {Phys. Rev. Lett.}\ }\textbf {\bibinfo {volume} {106}},\
  \bibinfo {pages} {207202} (\bibinfo {year} {2011})}\BibitemShut {NoStop}%
\bibitem [{\citenamefont {Albuquerque}\ \emph {et~al.}(2012)\citenamefont
  {Albuquerque}, \citenamefont {Alet},\ and\ \citenamefont
  {Moessner}}]{albuquerque12}%
  \BibitemOpen
  \bibfield  {author} {\bibinfo {author} {\bibfnamefont {A.~F.}\ \bibnamefont
  {Albuquerque}}, \bibinfo {author} {\bibfnamefont {F.}~\bibnamefont {Alet}}, \
  and\ \bibinfo {author} {\bibfnamefont {R.}~\bibnamefont {Moessner}},\ }\href
  {\doibase 10.1103/PhysRevLett.109.147204} {\bibfield  {journal} {\bibinfo
  {journal} {Phys. Rev. Lett.}\ }\textbf {\bibinfo {volume} {109}},\ \bibinfo
  {pages} {147204} (\bibinfo {year} {2012})}\BibitemShut {NoStop}%
\bibitem [{\citenamefont {Borzi}\ \emph {et~al.}(2013)\citenamefont {Borzi},
  \citenamefont {Slobinsky},\ and\ \citenamefont {Grigera}}]{borzi13}%
  \BibitemOpen
  \bibfield  {author} {\bibinfo {author} {\bibfnamefont {R.~A.}\ \bibnamefont
  {Borzi}}, \bibinfo {author} {\bibfnamefont {D.}~\bibnamefont {Slobinsky}}, \
  and\ \bibinfo {author} {\bibfnamefont {S.~A.}\ \bibnamefont {Grigera}},\
  }\href {\doibase 10.1103/PhysRevLett.111.147204} {\bibfield  {journal}
  {\bibinfo  {journal} {Phys. Rev. Lett.}\ }\textbf {\bibinfo {volume} {111}},\
  \bibinfo {pages} {147204} (\bibinfo {year} {2013})}\BibitemShut {NoStop}%
\bibitem [{\citenamefont {Guruciaga}\ \emph {et~al.}(2014)\citenamefont
  {Guruciaga}, \citenamefont {Grigera},\ and\ \citenamefont {Borzi}}]{borzi14}%
  \BibitemOpen
  \bibfield  {author} {\bibinfo {author} {\bibfnamefont {P.~C.}\ \bibnamefont
  {Guruciaga}}, \bibinfo {author} {\bibfnamefont {S.~A.}\ \bibnamefont
  {Grigera}}, \ and\ \bibinfo {author} {\bibfnamefont {R.~A.}\ \bibnamefont
  {Borzi}},\ }\href {\doibase 10.1103/PhysRevB.90.184423} {\bibfield  {journal}
  {\bibinfo  {journal} {Phys. Rev. B}\ }\textbf {\bibinfo {volume} {90}},\
  \bibinfo {pages} {184423} (\bibinfo {year} {2014})}\BibitemShut {NoStop}%
\bibitem [{\citenamefont {Brooks-Bartlett}\ \emph {et~al.}(2014)\citenamefont
  {Brooks-Bartlett}, \citenamefont {Banks}, \citenamefont {Jaubert},
  \citenamefont {Harman-Clarke},\ and\ \citenamefont {Holdsworth}}]{brooks14}%
  \BibitemOpen
  \bibfield  {author} {\bibinfo {author} {\bibfnamefont {M.~E.}\ \bibnamefont
  {Brooks-Bartlett}}, \bibinfo {author} {\bibfnamefont {S.~T.}\ \bibnamefont
  {Banks}}, \bibinfo {author} {\bibfnamefont {L.~D.~C.}\ \bibnamefont
  {Jaubert}}, \bibinfo {author} {\bibfnamefont {A.}~\bibnamefont
  {Harman-Clarke}}, \ and\ \bibinfo {author} {\bibfnamefont {P.~C.~W.}\
  \bibnamefont {Holdsworth}},\ }\href {\doibase 10.1103/PhysRevX.4.011007}
  {\bibfield  {journal} {\bibinfo  {journal} {Phys. Rev. X}\ }\textbf {\bibinfo
  {volume} {4}},\ \bibinfo {pages} {011007} (\bibinfo {year}
  {2014})}\BibitemShut {NoStop}%
\bibitem [{\citenamefont {Powell}(2015)}]{powell15}%
  \BibitemOpen
  \bibfield  {author} {\bibinfo {author} {\bibfnamefont {S.}~\bibnamefont
  {Powell}},\ }\href {\doibase 10.1103/PhysRevB.91.094431} {\bibfield
  {journal} {\bibinfo  {journal} {Phys. Rev. B}\ }\textbf {\bibinfo {volume}
  {91}},\ \bibinfo {pages} {094431} (\bibinfo {year} {2015})}\BibitemShut
  {NoStop}%
\bibitem [{\citenamefont {Jaubert}(2015)}]{jaubert15}%
  \BibitemOpen
  \bibfield  {author} {\bibinfo {author} {\bibfnamefont {L.~D.~C.}\
  \bibnamefont {Jaubert}},\ }\href {\doibase 10.1142/S2010324715400056}
  {\bibfield  {journal} {\bibinfo  {journal} {SPIN}\ }\textbf {\bibinfo
  {volume} {05}},\ \bibinfo {pages} {1540005} (\bibinfo {year}
  {2015})}\BibitemShut {NoStop}%
\bibitem [{\citenamefont {Bak}(1982)}]{bak82}%
  \BibitemOpen
  \bibfield  {author} {\bibinfo {author} {\bibfnamefont {P.}~\bibnamefont
  {Bak}},\ }\href {http://stacks.iop.org/0034-4885/45/i=6/a=001} {\bibfield
  {journal} {\bibinfo  {journal} {Reports on Progress in Physics}\ }\textbf
  {\bibinfo {volume} {45}},\ \bibinfo {pages} {587} (\bibinfo {year}
  {1982})}\BibitemShut {NoStop}%
\bibitem [{\citenamefont {Fennell}\ \emph {et~al.}(2005)\citenamefont
  {Fennell}, \citenamefont {Petrenko}, \citenamefont {F\aa{}k}, \citenamefont
  {Gardner}, \citenamefont {Bramwell},\ and\ \citenamefont
  {Ouladdiaf}}]{fennell05}%
  \BibitemOpen
  \bibfield  {author} {\bibinfo {author} {\bibfnamefont {T.}~\bibnamefont
  {Fennell}}, \bibinfo {author} {\bibfnamefont {O.~A.}\ \bibnamefont
  {Petrenko}}, \bibinfo {author} {\bibfnamefont {B.}~\bibnamefont {F\aa{}k}},
  \bibinfo {author} {\bibfnamefont {J.~S.}\ \bibnamefont {Gardner}}, \bibinfo
  {author} {\bibfnamefont {S.~T.}\ \bibnamefont {Bramwell}}, \ and\ \bibinfo
  {author} {\bibfnamefont {B.}~\bibnamefont {Ouladdiaf}},\ }\href {\doibase
  10.1103/PhysRevB.72.224411} {\bibfield  {journal} {\bibinfo  {journal} {Phys.
  Rev. B}\ }\textbf {\bibinfo {volume} {72}},\ \bibinfo {pages} {224411}
  (\bibinfo {year} {2005})}\BibitemShut {NoStop}%
\bibitem [{\citenamefont {Jaubert}\ \emph {et~al.}(2008)\citenamefont
  {Jaubert}, \citenamefont {Chalker}, \citenamefont {Holdsworth},\ and\
  \citenamefont {Moessner}}]{jaubert08}%
  \BibitemOpen
  \bibfield  {author} {\bibinfo {author} {\bibfnamefont {L.~D.~C.}\
  \bibnamefont {Jaubert}}, \bibinfo {author} {\bibfnamefont {J.~T.}\
  \bibnamefont {Chalker}}, \bibinfo {author} {\bibfnamefont {P.~C.~W.}\
  \bibnamefont {Holdsworth}}, \ and\ \bibinfo {author} {\bibfnamefont
  {R.}~\bibnamefont {Moessner}},\ }\href {\doibase
  10.1103/PhysRevLett.100.067207} {\bibfield  {journal} {\bibinfo  {journal}
  {Phys. Rev. Lett.}\ }\textbf {\bibinfo {volume} {100}},\ \bibinfo {pages}
  {067207} (\bibinfo {year} {2008})}\BibitemShut {NoStop}%
\bibitem [{\citenamefont {Powell}\ and\ \citenamefont
  {Chalker}(2008)}]{powell08}%
  \BibitemOpen
  \bibfield  {author} {\bibinfo {author} {\bibfnamefont {S.}~\bibnamefont
  {Powell}}\ and\ \bibinfo {author} {\bibfnamefont {J.~T.}\ \bibnamefont
  {Chalker}},\ }\href {\doibase 10.1103/PhysRevB.78.024422} {\bibfield
  {journal} {\bibinfo  {journal} {Phys. Rev. B}\ }\textbf {\bibinfo {volume}
  {78}},\ \bibinfo {pages} {024422} (\bibinfo {year} {2008})}\BibitemShut
  {NoStop}%
\bibitem [{\citenamefont {Blote}\ and\ \citenamefont
  {Hilborst}(1982)}]{blote82}%
  \BibitemOpen
  \bibfield  {author} {\bibinfo {author} {\bibfnamefont {H.~W.~J.}\
  \bibnamefont {Blote}}\ and\ \bibinfo {author} {\bibfnamefont {H.~J.}\
  \bibnamefont {Hilborst}},\ }\href
  {http://stacks.iop.org/0305-4470/15/i=11/a=011} {\bibfield  {journal}
  {\bibinfo  {journal} {Journal of Physics A: Mathematical and General}\
  }\textbf {\bibinfo {volume} {15}},\ \bibinfo {pages} {L631} (\bibinfo {year}
  {1982})}\BibitemShut {NoStop}%
\bibitem [{\citenamefont {Sylju\aa{}sen}\ and\ \citenamefont
  {Sandvik}(2002)}]{syljuasen02}%
  \BibitemOpen
  \bibfield  {author} {\bibinfo {author} {\bibfnamefont {O.~F.}\ \bibnamefont
  {Sylju\aa{}sen}}\ and\ \bibinfo {author} {\bibfnamefont {A.~W.}\ \bibnamefont
  {Sandvik}},\ }\href {\doibase 10.1103/PhysRevE.66.046701} {\bibfield
  {journal} {\bibinfo  {journal} {Phys. Rev. E}\ }\textbf {\bibinfo {volume}
  {66}},\ \bibinfo {pages} {046701} (\bibinfo {year} {2002})}\BibitemShut
  {NoStop}%
\bibitem [{\citenamefont {Sandvik}\ and\ \citenamefont
  {Moessner}(2006)}]{sandvik06}%
  \BibitemOpen
  \bibfield  {author} {\bibinfo {author} {\bibfnamefont {A.~W.}\ \bibnamefont
  {Sandvik}}\ and\ \bibinfo {author} {\bibfnamefont {R.}~\bibnamefont
  {Moessner}},\ }\href {\doibase 10.1103/PhysRevB.73.144504} {\bibfield
  {journal} {\bibinfo  {journal} {Phys. Rev. B}\ }\textbf {\bibinfo {volume}
  {73}},\ \bibinfo {pages} {144504} (\bibinfo {year} {2006})}\BibitemShut
  {NoStop}%
\bibitem [{\citenamefont {Alet}\ \emph {et~al.}(2006)\citenamefont {Alet},
  \citenamefont {Ikhlef}, \citenamefont {Jacobsen}, \citenamefont {Misguich},\
  and\ \citenamefont {Pasquier}}]{alet06}%
  \BibitemOpen
  \bibfield  {author} {\bibinfo {author} {\bibfnamefont {F.}~\bibnamefont
  {Alet}}, \bibinfo {author} {\bibfnamefont {Y.}~\bibnamefont {Ikhlef}},
  \bibinfo {author} {\bibfnamefont {J.~L.}\ \bibnamefont {Jacobsen}}, \bibinfo
  {author} {\bibfnamefont {G.}~\bibnamefont {Misguich}}, \ and\ \bibinfo
  {author} {\bibfnamefont {V.}~\bibnamefont {Pasquier}},\ }\href {\doibase
  10.1103/PhysRevE.74.041124} {\bibfield  {journal} {\bibinfo  {journal} {Phys.
  Rev. E}\ }\textbf {\bibinfo {volume} {74}},\ \bibinfo {pages} {041124}
  (\bibinfo {year} {2006})}\BibitemShut {NoStop}%
\bibitem [{\citenamefont {Glosli}\ and\ \citenamefont
  {Plischke}(1983)}]{glosli83}%
  \BibitemOpen
  \bibfield  {author} {\bibinfo {author} {\bibfnamefont {J.}~\bibnamefont
  {Glosli}}\ and\ \bibinfo {author} {\bibfnamefont {M.}~\bibnamefont
  {Plischke}},\ }\href {\doibase 10.1139/p83-197} {\bibfield  {journal}
  {\bibinfo  {journal} {Canadian Journal of Physics}\ }\textbf {\bibinfo
  {volume} {61}},\ \bibinfo {pages} {1515} (\bibinfo {year}
  {1983})}\BibitemShut {NoStop}%
\bibitem [{\citenamefont {Takagi}\ and\ \citenamefont
  {Mekata}(1995)}]{takagi95}%
  \BibitemOpen
  \bibfield  {author} {\bibinfo {author} {\bibfnamefont {T.}~\bibnamefont
  {Takagi}}\ and\ \bibinfo {author} {\bibfnamefont {M.}~\bibnamefont
  {Mekata}},\ }\href {\doibase 10.1143/JPSJ.64.4609} {\bibfield  {journal}
  {\bibinfo  {journal} {Journal of the Physical Society of Japan}\ }\textbf
  {\bibinfo {volume} {64}},\ \bibinfo {pages} {4609} (\bibinfo {year}
  {1995})}\BibitemShut {NoStop}%
\bibitem [{\citenamefont {Rastelli}\ \emph {et~al.}(2005)\citenamefont
  {Rastelli}, \citenamefont {Regina},\ and\ \citenamefont
  {Tassi}}]{rastelli05}%
  \BibitemOpen
  \bibfield  {author} {\bibinfo {author} {\bibfnamefont {E.}~\bibnamefont
  {Rastelli}}, \bibinfo {author} {\bibfnamefont {S.}~\bibnamefont {Regina}}, \
  and\ \bibinfo {author} {\bibfnamefont {A.}~\bibnamefont {Tassi}},\ }\href
  {\doibase 10.1103/PhysRevB.71.174406} {\bibfield  {journal} {\bibinfo
  {journal} {Phys. Rev. B}\ }\textbf {\bibinfo {volume} {71}},\ \bibinfo
  {pages} {174406} (\bibinfo {year} {2005})}\BibitemShut {NoStop}%
\bibitem [{\citenamefont {Yokoi}\ \emph {et~al.}()\citenamefont {Yokoi},
  \citenamefont {Nagle},\ and\ \citenamefont {Salinas}}]{yokoi86}%
  \BibitemOpen
  \bibfield  {author} {\bibinfo {author} {\bibfnamefont {C.~S.~O.}\
  \bibnamefont {Yokoi}}, \bibinfo {author} {\bibfnamefont {J.~F.}\ \bibnamefont
  {Nagle}}, \ and\ \bibinfo {author} {\bibfnamefont {S.~R.}\ \bibnamefont
  {Salinas}},\ }\href {\doibase 10.1007/BF01011905} {\bibfield  {journal}
  {\bibinfo  {journal} {Journal of Statistical Physics}\ }\textbf {\bibinfo
  {volume} {44}},\ \bibinfo {pages} {729}}\BibitemShut {NoStop}%
\bibitem [{\citenamefont {Jiang}\ and\ \citenamefont {Emig}(2006)}]{jiang06}%
  \BibitemOpen
  \bibfield  {author} {\bibinfo {author} {\bibfnamefont {Y.}~\bibnamefont
  {Jiang}}\ and\ \bibinfo {author} {\bibfnamefont {T.}~\bibnamefont {Emig}},\
  }\href {\doibase 10.1103/PhysRevB.73.104452} {\bibfield  {journal} {\bibinfo
  {journal} {Phys. Rev. B}\ }\textbf {\bibinfo {volume} {73}},\ \bibinfo
  {pages} {104452} (\bibinfo {year} {2006})}\BibitemShut {NoStop}%
\bibitem [{\citenamefont {Powell}\ and\ \citenamefont
  {Chalker}(2009)}]{powell09}%
  \BibitemOpen
  \bibfield  {author} {\bibinfo {author} {\bibfnamefont {S.}~\bibnamefont
  {Powell}}\ and\ \bibinfo {author} {\bibfnamefont {J.~T.}\ \bibnamefont
  {Chalker}},\ }\href {\doibase 10.1103/PhysRevB.80.134413} {\bibfield
  {journal} {\bibinfo  {journal} {Phys. Rev. B}\ }\textbf {\bibinfo {volume}
  {80}},\ \bibinfo {pages} {134413} (\bibinfo {year} {2009})}\BibitemShut
  {NoStop}%
\end{thebibliography}%

%%%%%%%%%%%%%%%%%%%%%%%%%%%%%%%%%%%%%%%%%%%%%%%%%%%%%%
\onecolumngrid
\appendix
%%%%%%%%%%%%%%%%%%%%%%%%%%%%%%%%%%%%%%%%%%%%%%%%%%%%%%

%%%%%%%%%%%%%%%%%%%%%%%%%%%%%%%%%%%%%%%%%%%%%%%%%%%%%%
% Winding number - dominance of (0,0) sector + triangle of possible values
\section{Winding number sectors}
\label{app:windsec}
%%%%%%%%%%%%%%%%%%%%%%%%%%%%%%%%%%%%%%%%%%%%%%%%%%%%%%

The ground state configurations of the triangular lattice Ising antiferromagnet (TLIAF) are highly degenerate, with a degeneracy $e^{0.323N}$, where $N$ is the number of lattice sites\cite{wannier50}.
These configurations can be labelled by a pair of winding numbers, as explained in the main text.
Here we explore the allowed values of the winding numbers, as well as determining the degeneracy of some of the important sectors.

\begin{figure}[h]
\centering
\includegraphics[width=0.75\textwidth]{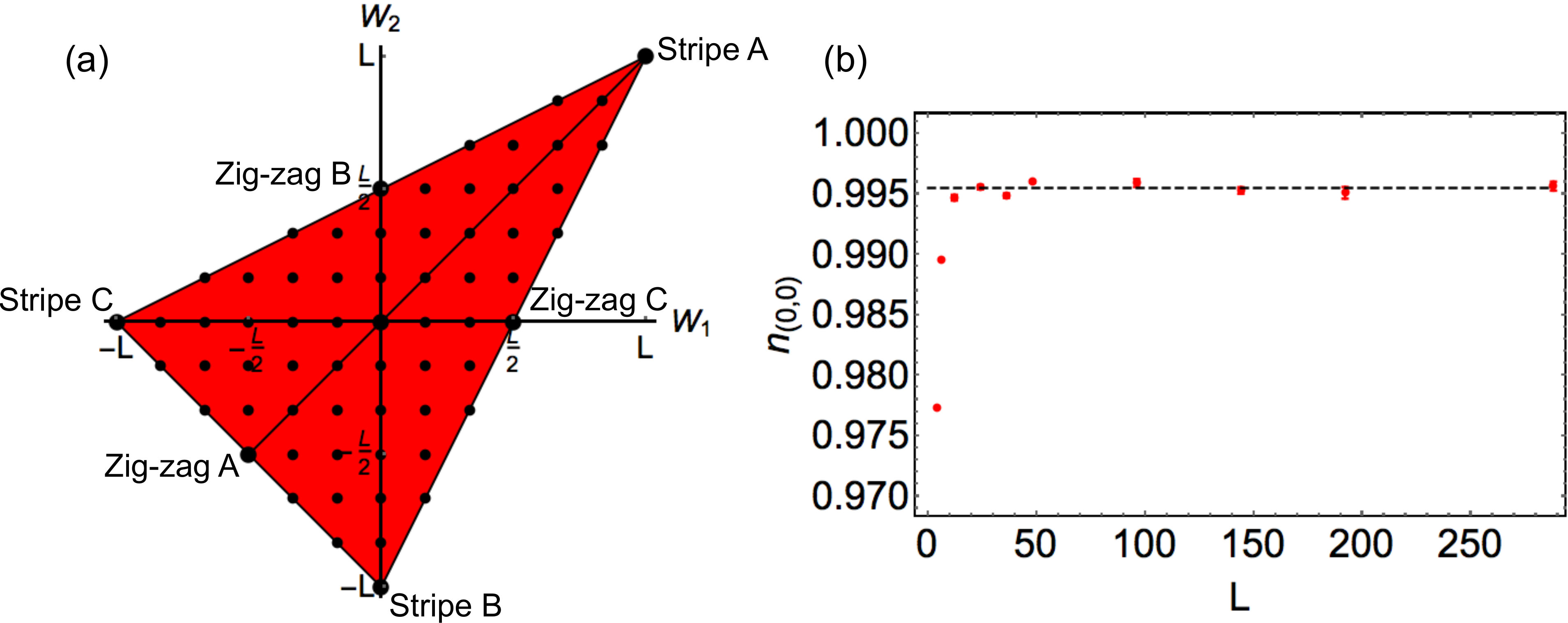}
\caption{\footnotesize{
Winding number sectors for the ground states of the nearest-neighbour triangular lattice Ising antiferromagnet.
(a) The winding numbers $W_1$ and $W_2$ are constrained to be multiples of 2 and contained within a triangle with vertices at $W=(L,L)$, $(-L,0)$ and $(0,-L)$.
The allowed winding number sectors are shown in the case of $L=12$.
Vertices of the triangle are two-fold degenerate and correspond to stripe configurations.
The addition of pairs of double domain walls moves the system along the line joining the corresponding vertex and the $(0,0)$ sector.
(b) The (0,0) sector has an extensive degeneracy, and Monte Carlo simulation shows that in the thermodynamic limit this corresponds to 0.996 of the total degeneracy.
Therefore the number of configurations in this sector is given by $0.996 e^{0.323N} = e^{0.319N}$.
}}
\label{fig:windsectors}
\end{figure}

A pair of winding numbers, $W=(W_1,W_2)$, are defined in terms of the dimers crossing two reference lines, as shown in Fig.~2 in the main text.
For hexagonal clusters with $N=3L^2$ sites and periodic boundary conditions the winding numbers are required to be multiples of 2 and to be contained by a triangle with vertices at $W=(L,L)$, $(-L,0)$ and $(0,-L)$, as shown in Fig.~\ref{fig:windsectors}.
Each of these vertices contains a pair of stripe configurations, with a fixed stripe direction and a two-fold degeneracy due to a global spin flip.

The zig-zag states are located in the winding number sectors $W=(L/2,0)$, $W=(-L/2,-L/2)$ and $W=(0,L/2)$ (assuming $L$ is a multiple of 4).
Each sector is four-fold degenerate and consists of zig-zags running in a fixed direction, with a global spin-flip degeneracy and a 2-fold translational degeneracy.

The addition of a pair of double domain walls to the stripe state results in a change of winding number sector that moves the system one position along the line joining the relevant vertex and the sector $(0,0)$.
The degeneracy of these sectors is proportional to $e^ L$ and is therefore large but sub-extensive.
The addition of further pairs of double domain walls moves the system in the direction of the $(0,0)$ sector, and also increases the degeneracy.

The sector $(0,0)$ has an extensive degeneracy, and contains the majority of the ground state configurations.
Monte Carlo simulations of the nearest neighbour model with a maximum size of $L=288$ show that in the thermodynamic limit this sector contains 0.996 of the configurations (see Fig.~\ref{fig:windsectors}).
Thus the degeneracy is \mbox{$0.996 e^{0.323N} = e^{0.319N}$}.

%%%%%%%%%%%%%%%%%%%%%%%%%%%%%%%%%%%%%%%%%%%%%%%%%%%%%%
% Monte Carlo Worm algorithm, taking into account J1-J5 interactions
\section{Worm algorithm for Monte Carlo updates}
\label{app:wormalg}
%%%%%%%%%%%%%%%%%%%%%%%%%%%%%%%%%%%%%%%%%%%%%%%%%%%%%%

In order to perform Monte Carlo simulations of the triangular lattice Ising antiferromagnet (TLIAF) with $J_1 \to \infty$ we have employed a worm algorithm\cite{syljuasen02,sandvik06,alet06}.
This allows the system to make transitions between different winding number sectors and is designed such that proposed Monte Carlo updates are always accepted.
We here describe in detail how such an algorithm can be built for the TLIAF.

\begin{figure}[t]
\centering
\includegraphics[width=0.7\textwidth]{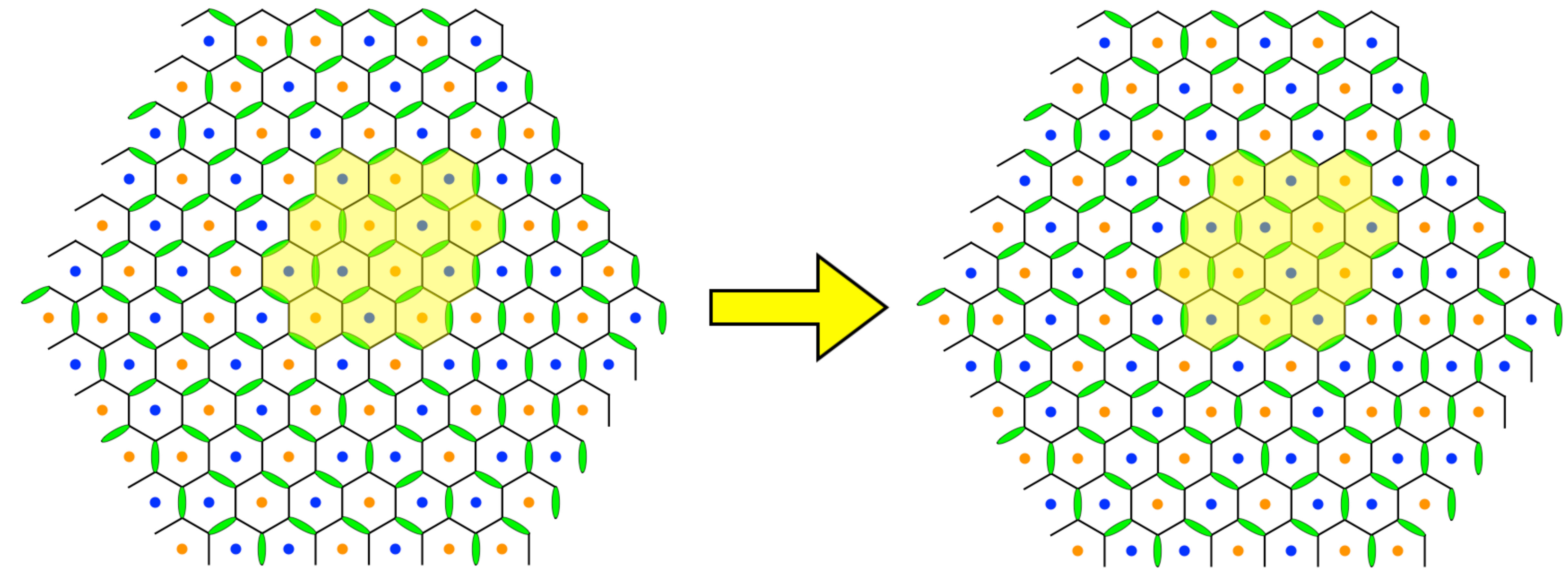}
\caption{\footnotesize{
Loop updates on the honeycomb lattice.
The ground states of the nearest-neighbour triangular lattice antiferromagnet can be mapped onto a dimer model on the honeycomb lattice (see Fig.~1 in the main text).
By forming loops of alternating dimer and empty bonds (edge of yellow region) and exchanging the two, a different ground state can be generated.
This is equivalent to flipping all the spins contained within the loop (yellow region).
This procedure, directed by further-neighbour interactions, forms the basis of the Monte Carlo updates used in this work.
}}
\label{fig:dimerloops}
\end{figure}

The creation of the worm algorithm starts from the mapping of the ground states of the nearest-neighbour TLIAF onto a dimer model on the honeycomb lattice (see Fig.~\ref{fig:dimerloops}).
%
%Alternatively this can be viewed as a fully packed loop model.
%
Starting from one particular dimer covering of the honeycomb lattice, there is a simple rule which allows other dimer coverings to be generated\cite{zhang09}.
This involves identifying closed loops of alternating dimer-covered and empty bonds and swapping the two.
In terms of the Ising model on the triangular lattice, this is equivalent to flipping all the spins enclosed within the loop, and an example is shown in Fig.~\ref{fig:dimerloops}.
In the presence of periodic boundary conditions there is the complication that only dimer configurations with even winding numbers, $W_1$ and $W_2$, are physical.
In consequence it is necesary to continue to perform loop updates until this condition is satisfied.
It has been shown that a Markov chain of states created in this way is ergodic\cite{zhang09}.

The addition of further-neighbour interactions partially lifts the degeneracy between ground states of the nearest-neighbour TLIAF. 
The most obvious way to take this into account in a Monte Carlo simulation is to propose loop updates on the honeycomb lattice, thus restricting the system to ground states of the nearest-neighbour model, and include a metropolis-like accept/reject step that takes into account the $J_2,J_3\dots$ interactions.
However, we find that at low temperature such an algorithm fails to equilibriate in a reasonable time, due to the gapped nature of the stripe ground state and the large degeneracy of the manifold.

The solution is to direct the formation of the loops on the honeycomb lattice using the further-neighbour interactions.
In order to do this we follow in spirit Ref.~[\onlinecite{syljuasen02,sandvik06,alet06}] in forming a directed loop or ``worm'' algorithm.
First we will consider guiding the loop formation using the $J_2$ and $J_3$ interactions, and then later we will briefly mention how to also include the $J_4$ and $J_5$ interactions.
Interactions beyond $J_5$ can in principle be included, but at an ever increasing computational cost.

\begin{figure}[h]
\centering
\includegraphics[width=0.9\textwidth]{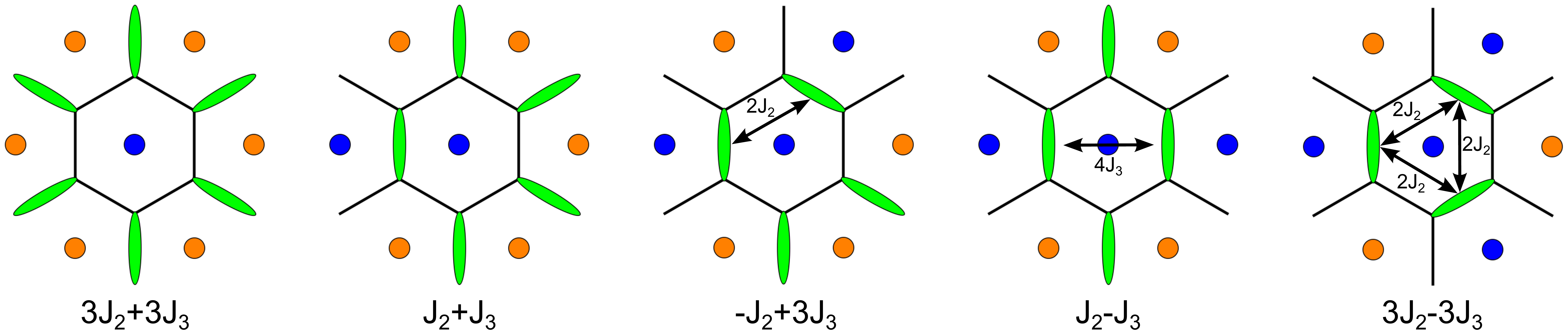}
\caption{\footnotesize{
Hexagon energies due to $J_2$ and $J_3$ interactions.
Hexagons are considered to have a half share of the six $J_2$ bonds and a full share of the  three $J_3$ bonds crossing them, thus defining a $J_2$-$J_3$ hexagon energy.
This can be recast in terms of dimer energies.
Each hexagon is given an initial energy of $3J_2+3J_3$.
For each edge covered by a dimer an energy of $-2J_2-2J_3$ is added.
Each second-neighbour dimer-dimer interaction on the hexagon costs an energy $2J_2$, while each third-neighbour dimer-dimer interactions costs an energy $4J_3$.
All distinct hexagon configurations are shown, along with their associated energies.
}}
\label{fig:hexagonenergies}
\end{figure}

The $J_2$ and $J_3$ interactions between Ising spins on the triangular lattice can be mapped onto interactions between dimers on the honeycomb lattice (see Fig.~\ref{fig:hexagonenergies}).
One way to do this is to consider the set of $N$ hexagons that make up the honeycomb lattice (there are $N$ sites on the triangular lattice).
Each hexagon is considered to have a half share of the six $J_2$ bonds crossing it and a full share of the three $J_3$ bonds, and can therefore be assigned a $J_2-J_3$ energy.
First an energy $3J_2+3J_3$ is assigned to every hexagon and then an energy $-2J_2-2J_3$ is added for each dimer covering one of the edges.
Pairs of dimers separated by a single intermediate uncovered bond are assigned the interaction energy $2J_2$, while dimers on opposite sides of the hexagon are given an interaction energy $4J_3$.
It can be checked that this correctly reproduces the Ising energies associated with the hexagon for all possible configurations. 

The aim is to create loops on the honeycomb lattice in such a way that the initial and final states are in detailed balance, and therefore the update is always accepted.
As a starting point a site on the honeycomb lattice, labelled $i_1$, is chosen at random, and there is guaranteed to be exactly one dimer emanating from it.
The head of the worm is moved from $i_1$ to the opposite end of the dimer, labelled $i_2$, and then there are two possibilities:
1) the head of the worm returns to $i_1$ (backtracks), the dimer remains where it is and the worm terminates since it has returned to the starting point;
2) the worm head moves to $i_3$, one of the two neighbours of $i_2$ that is not $i_1$, the dimer on the bond $i_1-i_2$ is deleted and a dimer is placed on the bond $i_2-i_3$.
In case 2) the system is in an unphysical state in which site $i_1$ is not covered by a dimer and site $i_3$ is covered by two dimers.
The head of the worm continues to move in the same manner until it returns to $i_1$ and closes the loop creating a new physical state (assuming the winding numbers remain even).

\begin{figure}[h]
\centering
\includegraphics[width=0.9\textwidth]{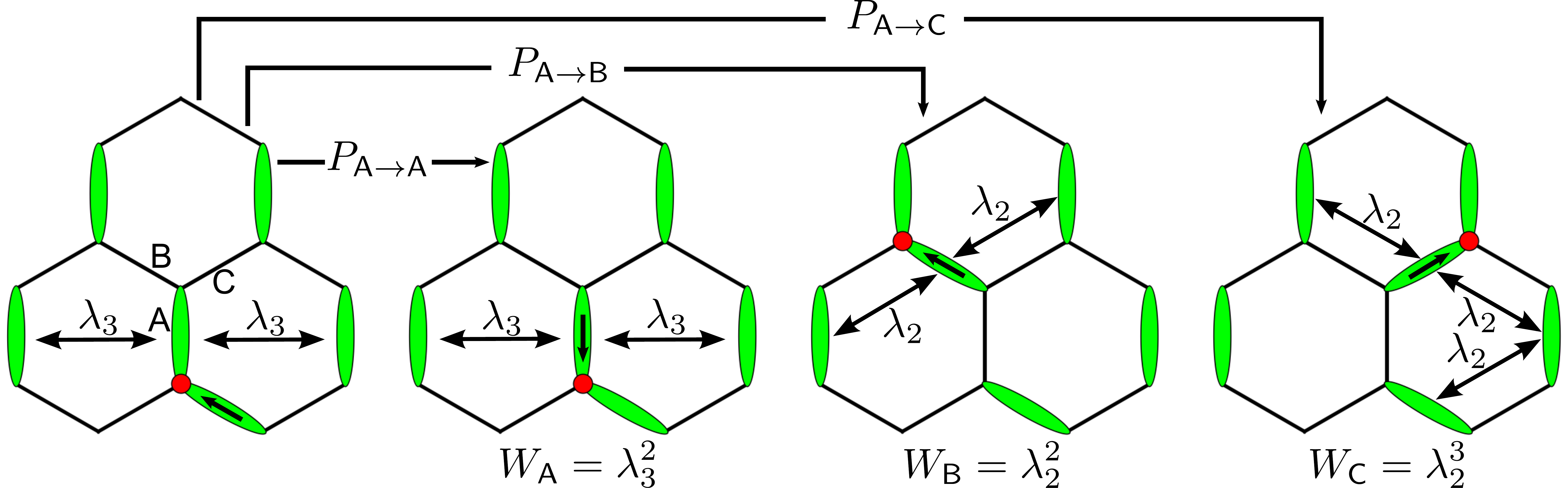}
\caption{\footnotesize{
An example of how to construct a set of move probabilities for the worm algorithm. 
The head of the worm is shown as a red dot, and the most recent bond traversed by a black arrow.
At all intermediate steps in the worm construction, the state of the system is unphysical, since there is a site covered with 2 dimers (green ellipses).
In the initial state (left hand side) there is a dimer on the central {\sf A} bond.
This dimer can either be moved to the {\sf B} or {\sf C} bond, or can remain on the {\sf A} bond while the worm reverses direction (backtracking).
A set of probabilites $P_{j \to k}$ can be determined for these moves by taking into account the relative weights, $W_j$ and $W_k$, of these states.
Worms constructed in this way obey detailed balance globally, and are therefore always accepted as an update.
}}
\label{fig:wormcreation}
\end{figure}

In order to achieve detailed balance it is necessary to carefully choose the probabilities that determine in which direction the head of the worm moves.
These probabilites depend on the local environment and are required to fulfill a number of conditions.
We define $P_{j\to k}$ where \mbox{$j,k \in \{ {\sf A}, {\sf B}, {\sf C} \} $}, and for example $P_{\sf A \to B}$ describes a situation in which the worm head traverses first an {\sf A} bond and then a {\sf B} bond, resulting in the deletion of a dimer on the {\sf A} bond and the creation of a dimer on the {\sf B} bond (see Fig.~\ref{fig:wormcreation}).
Similarly $P_{\sf A \to A}$ describes the worm head traversing the {\sf A} bond first in one direction and then in the opposite direction, reversing the direction of travel of the worm, but not changing the dimer configuration.
This probability can be written as,
\begin{align}
 P_{j \to k} = \frac{a_{j\to k}}{W_j},
 \label{eq:P=a/W}
\end{align}
where $W_j$ is the statistical weight associated with having a dimer on bond $j$.
This weight arises from the interactions with other nearby dimers, and is given by \mbox{$W_j=\exp[-E_{{\sf int},j}/T]$}, where $E_{{\sf int},j}$ is the total interaction energy associated with there being a dimer on bond $j$. 
The conditions are that the probabilities sum to 1,
\begin{align}
\sum_{k} P_{j \to k} =1,
\label{eq:P=1}
\end{align}
and that detailed balance is obeyed at every step of the worm creation,
\begin{align}
a_{j \to k} = a_{k \to j}.
\label{eq:detbalance}
\end{align}
By locally observing detailed balance it is guaranteed that the worm update as a whole obeys detailed balance.

Combining the conditions described by Eq.~\ref{eq:P=1} and Eq.~\ref{eq:detbalance} lead to the weight equations,
\begin{align}
W_{\sf A} = a_{\sf A \to A} + a_{\sf A \to B} + a _{\sf C \to A}, \quad
W_{\sf B} = a_{\sf B \to B} + a_{\sf B \to C} + a _{\sf A \to B}, \quad
W_{\sf C} = a_{\sf C \to C} + a_{\sf C \to A} + a _{\sf B \to C},
\label{eq:weights}
\end{align}
and it can be seen that these equations are underconstrained, since there are 3 equations and 6 unknowns.
We choose to minimise the backtracking terms $a_{j\to j}$, in the belief that this should make the algorithm more efficient.
When possible we set $a_{j\to j}=0$, but this is not always allowed.

Since the weights depend on the local dimer configuration of the three hexagons surrounding a vertex, it is necessary to solve the weight equations, Eq.~\ref{eq:weights}, for each possible local configuration (or at least each possible class).
This is done numerically, and here we just show one key example, where the local configuration is shown in Fig.~\ref{fig:wormcreation}.
The weights are given by,
\begin{align}
W_{\sf A} = \lambda_3^2, \quad
W_{\sf B} = \lambda_2^2, \quad
W_{\sf C} = \lambda_2^3 ,
\end{align}
where $\lambda_2 = \exp[-2J_2/T]$ and $\lambda_3=\exp[-4J_3/T]$.
Let us first consider solving the weight equations [Eq.~\ref{eq:weights}] with $a_{j \to j}=0$ (i.e. no backtracking).
In this case there are as many equations as unknowns and one finds,
\begin{align}
a_{\sf A \to B} &= a_{\sf B \to A} = 
\frac{1}{2} \left( W_{\sf A} + W_{\sf B} - W_{\sf C} \right) =
\frac{1}{2} \left( \lambda_3^2 +  \lambda_2^2 - \lambda_2^3 \right)   \nonumber \\
a_{\sf B \to C} &= a_{\sf C \to B} =
\frac{1}{2} \left( -W_{\sf A} + W_{\sf B} + W_{\sf C} \right) =
\frac{1}{2} \left( -\lambda_3^2 +  \lambda_2^2 + \lambda_2^3 \right)   \nonumber \\
a_{\sf C \to A} &= a_{\sf A \to C} = 
\frac{1}{2} \left( W_{\sf A} - W_{\sf B} + W_{\sf C} \right) =
\frac{1}{2} \left( \lambda_3^2 -  \lambda_2^2 + \lambda_2^3 \right)   ,
\label{eq:nobounce}
\end{align}
and the transition probabilities follow from Eq.~\ref{eq:P=a/W}.
At temperatures high compared to $J_2$ and $J_3$  then $\lambda_2,\lambda_3 \to 1$, and there is an equal probability of moving the dimer to either of the unvisited bonds, with zero probability of backtracking.

As temperature is lowered, the solution shown in Eq.~\ref{eq:nobounce} remains valid if all the transition probabilites are positive.
If we consider the case where $J_3 < J_2/2$ (i.e. there is a stripe ground state), then the first transition element to become negative is $a_{\sf B \to C} $, and the temperature at which this occurs is found by solving the equation,
\begin{align}
 -\lambda_3^2 +  \lambda_2^2 + \lambda_2^3 = 0.
\end{align}
It is interesting to note that this is the same equation as is solved in Ref.~[\onlinecite{korshunov05}] to find the exact temperature, $T_{\sf ddw}$, at which double domain walls appear.

For $T<T_{\sf ddw}$ it is no longer possible to solve the weight equations without including backtracking, and we choose to minimise the backtracking term.
For temperatures below $T_{\sf ddw}$, it is sufficient set $a_{\sf A \to A} \neq 0$ while $a_{\sf B \to B} = a_{\sf C \to C} = 0$.
We choose the solution,
\begin{align}
a_{\sf A \to A} &= \lambda_3^2 -  \lambda_2^2 - \lambda_2^3 \nonumber \\
a_{\sf A \to B} &= a_{\sf B \to A} = 
\frac{1}{2} \left( \lambda_3^2 +  \lambda_2^2 - \lambda_2^3 \right) - \frac{1}{2} \left( \lambda_3^2 -  \lambda_2^2 - \lambda_2^3 \right)  = \lambda_2^2 \nonumber \\
a_{\sf B \to C} &= a_{\sf C \to B} =
0   \nonumber \\
a_{\sf C \to A} &= a_{\sf A \to C} = 
\frac{1}{2} \left( \lambda_3^2 -  \lambda_2^2 + \lambda_2^3 \right) -   \frac{1}{2} \left( \lambda_3^2 -  \lambda_2^2 - \lambda_2^3 \right)  =  \lambda_2^3.
\label{eq:bounce}
\end{align}
It can be seen that as $T \to T_{\sf ddw}^-$ this matches the no-backtracking solution given in Eq.~\ref{eq:nobounce}, and the probabilities thus vary smoothly with temperature.
A similar procedure can be implemented numerically to determine the probabilities for all other possible local environments.

The inclusion of $J_4$ and $J_5$ interactions into the worm algorithm is conceptually similar.
$J_4$ and $J_5$ interactions can be mapped onto interactions between 3 bonds (dimer or empty) on the honeycomb lattice.
The weight equations [Eq.~\ref{eq:weights}] can be solved numerically, taking into account an expanded local environment.

When performing simulations we typically perform $10^5$ worm updates per temperature for thermalisation, with a parallel tempering step after each 10 worm updates.
This is then followed by $10^5$ worm updates per temperature for measurement, again with a parallel tempering step after each 10 worm updates.
%

%%%%%%%%%%%%%%%%%%%%%%%%%%%%%%%%%%%%%%%%%%%%%%%%%%%%%%
% Simulations of J1-J2-J3 model
\section{Triangular lattice Ising antiferromagnet with J$_2$ and J$_3$ interactions}
\label{app:J1J2J3}
%%%%%%%%%%%%%%%%%%%%%%%%%%%%%%%%%%%%%%%%%%%%%%%%%%%%%%

We here provide a more detailed study of the $J_1$-$J_2$-$J_3$ triangular lattice Ising antiferromagnet with $J_1\to \infty$.
We use Monte Carlo simulation to study the phase diagram and the structure factor, and confirm numerically many of the predictions made in Ref.~[\onlinecite{korshunov05}].
We note that Monte Carlo simulations of the $J_1$-$J_2$ model have previously been carried out in Ref.~[\onlinecite{glosli83,takagi95,rastelli05}], but these studies were not able to access the regime $J_1 \gg J_2$ due to the use of local updates.

\begin{figure}[h]
\centering
\includegraphics[width=0.9\textwidth]{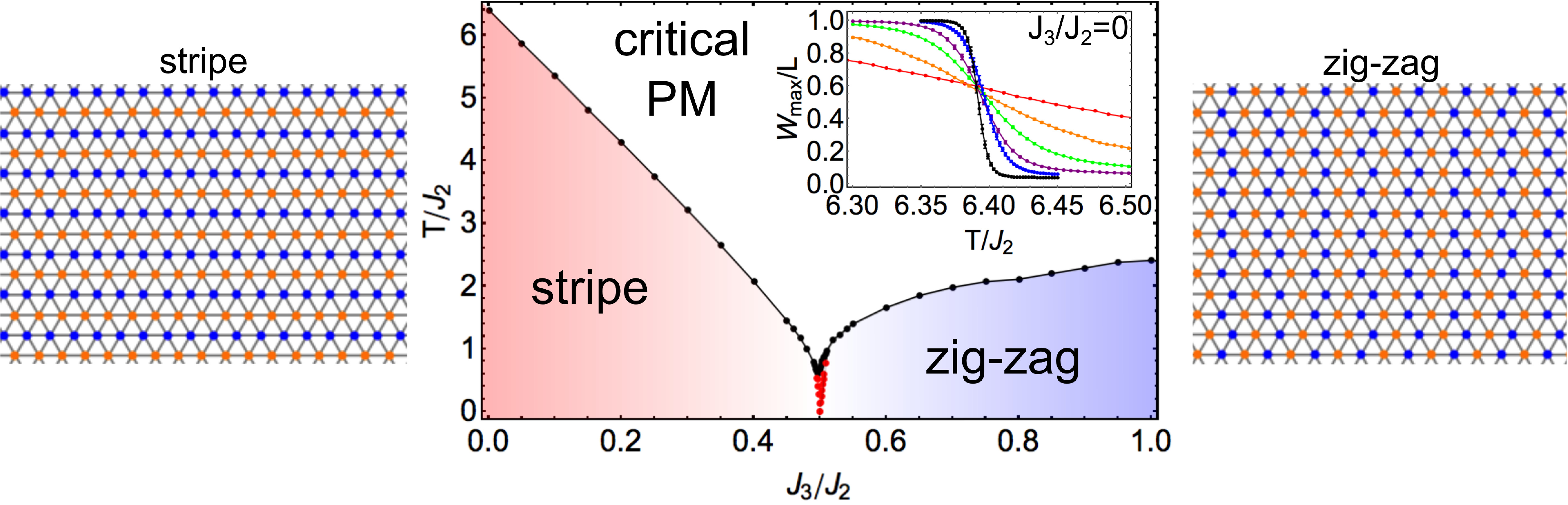}
\caption{\footnotesize{
Phase diagram of the $J_1$-$J_2$-$J_3$ triangular lattice antiferromagnetic Ising model with $J_1 \to \infty$, as determined by Monte Carlo simulation.
For $J_3/J_2<0.5$ the ground state consists of stripes of equivalent Ising spins (shown on left) while for $J_3/J_2>0.5$ it consists of zig-zags (shown on right).
The point $J_3/J_2=0.5$ has a large but sub-extensive degeneracy, since straight domain walls can be placed into the stripe state at zero energy cost.
The black points show the location of 1st order phase transitions, as determined by Monte Carlo simulations with $L=36$.
The red points enclose an additional phase, and the region of parameter space occupied by this phase can be shown analytically to tend to zero in the thermodynamic limit.
The inset shows the winding number $W_{\sf max} = \mathrm{max}(|W_1|,|W_2|,|W_2-W_1|)$ for the model with $J_3/J_2=0$ in the vicinity of the phase transition.
Cluster sizes are $L=12$ (red), $L=18$ (orange), $L=24$ (green), $L=30$ (purple), $L=36$ (blue) and $L=48$ (black).
}}
\label{fig:J2J3phasediag}
\end{figure}

The phase diagram is shown in Fig.~\ref{fig:J2J3phasediag}.
The ground state for $J_3/J_2<0.5$ is the stripe state and for $J_3/J_2>0.5$ is the zig-zag state.
On increasing the temperature these both have a first-order phase transition to the critical paramagnet.
In the case $J_3=0$ we find a critical temperature \mbox{$T_{1} = 6.39 J_2$}.

At $J_3/J_2=0.5$ there is a high-degeneracy point, in which any number of straight, parallel domain walls can be placed into the stripe state (the zig-zag state corresponds to a fully packed set of straight domain walls). 
This corresponds to a degeneracy between all the winding number sectors lying on the edges of the triangle shown in Fig.~\ref{fig:windsectors}a).
The degeneracy is proportional to $2^L$ and is therefore large but sub-extensive.

One question that can be asked is whether thermal fluctuations can entropically stabilise a state with a fluctuating set of straight domain walls away from $J_3/J_2=0.5$.
In finite-size Monte Carlo simulations we find that this is the case, and the region of stability is shown by red dots in Fig.~\ref{fig:J2J3phasediag}.
However, it can be shown that in the thermodynamic limit the $J_3/J_2$ width of this phase tends to zero.

In order to do this it is useful to define a quantity $\Delta$ that measures the distance in parameter space from the high degeneracy point, according to $J_3=J_2/2-\Delta$.
The energy cost per unit length of a domain wall is therefore given by $E_{\sf dw} = 4\Delta$ and straight domain walls placed into hexagonal clusters have length $3L$.
The partition function, with energy measured relative to the stripe state, is given by,
\begin{align}
\mathcal{Z}_{\sf sdw}  = \sum_{k=1}^L  \frac{(L)!}{k! (L-k)!} e^{-12k L\Delta /T} 
= \left( 1+e^{-12 L\Delta /T} \right)^L -1,
\end{align}
where sdw stands for straight domain wall and $k$ is the number of domain walls in the system.
The corresponding free energy is given by,
\begin{align}
\mathcal{F}_{\sf sdw}(T)  =-T \log \left[ \left( 1+e^{-12 L\Delta /T} \right)^L -1 \right],
\end{align}
and the sdw state is stable when this is negative.
The boundary between the stripe and sdw phase is determined from solving $\mathcal{F}_{\sf sdw}(T_{\sf sdw})=0$, and this gives,
\begin{align}
T_{\sf sdw} = -\frac{12 L \Delta}{\log \left[ 2^{1/L} -1 \right]}.
\label{eq:Tsdw}
\end{align}
It can be seen that for any finite value of $\Delta$ then  $T_{\sf sdw} \to \infty$ as $L \to \infty$, and therefore the sdw phase disappears in the thermodynamic limit.
A completely analagous logic can be used in the zig-zag state.
Simulations on finite lattices confirm this behaviour, and the transition temperature to the sdw state matches very well with the analytically calculated value in Eq.~\ref{eq:Tsdw}. 

\begin{figure}[h]
\centering
\includegraphics[width=0.99\textwidth]{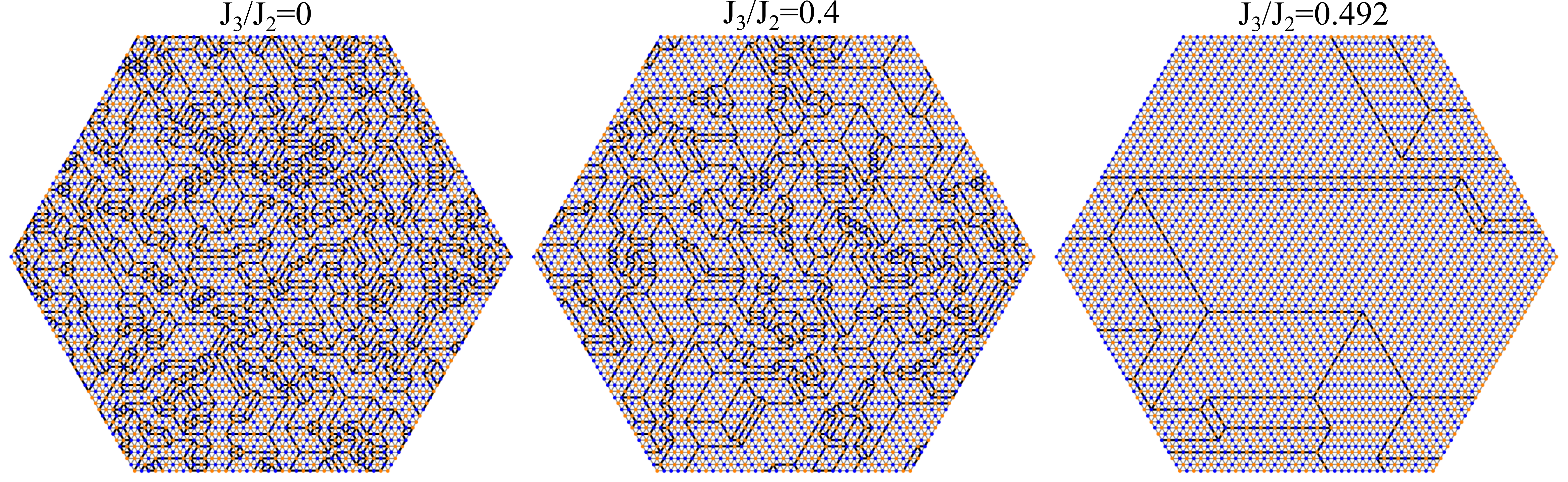}
\caption{\footnotesize{
Representative Monte Carlo simulation snapshots of the critical paramagnet just above the first order phase transition (see Fig.~\ref{fig:J2J3phasediag}).
Black lines separate domains with stripes running in different directions.
The typical domain size is small at $J_3/J_2=0$ and increases approaching the high degeneracy point at $J_3/J_2=0.5$.
}}
\label{fig:PMsnapshots}
\end{figure}

The phase transition between the stripe state and the critical paramagnet is first order, and this has been confirmed by  the observation of a double peak structure in the energy histogram close to the transition temperature.
As $J_3/J_2$ is increased towards the high degeneracy point the transition becomes progressively less first order, in the sense that the splitting of the two peaks in the energy histogram reduces.
As explained above, Monte Carlo simulations are restricted as to how closely the high degeneracy point can be approached, but it appears that the energy gap scales to zero at this point, and therefore $J_3/J_2=0.5$ is a $T=0$ critical endpoint of a line of 1st order transitions.
The physics is clearer if one studies a set of representative snapshots of the critical paramagnet just above the transition (Fig.~\ref{fig:PMsnapshots}).
At $J_3/J_2=0$ the system forms very small domains of stripe order, and therefore there is a large energy jump associated with the transition.
Increasing $J_3/J_2$ leads to larger domain sizes just above the transition, and therefore a smaller energy jump.
Approaching $J_3/J_2=0.5$ the domain size appears to diverge, leading to a $T=0$ critical endpoint where the energy jump tends to zero.
This is in agreement with analysis of a height model\cite{korshunov05}.

\begin{figure}[h]
\centering
\includegraphics[width=0.75\textwidth]{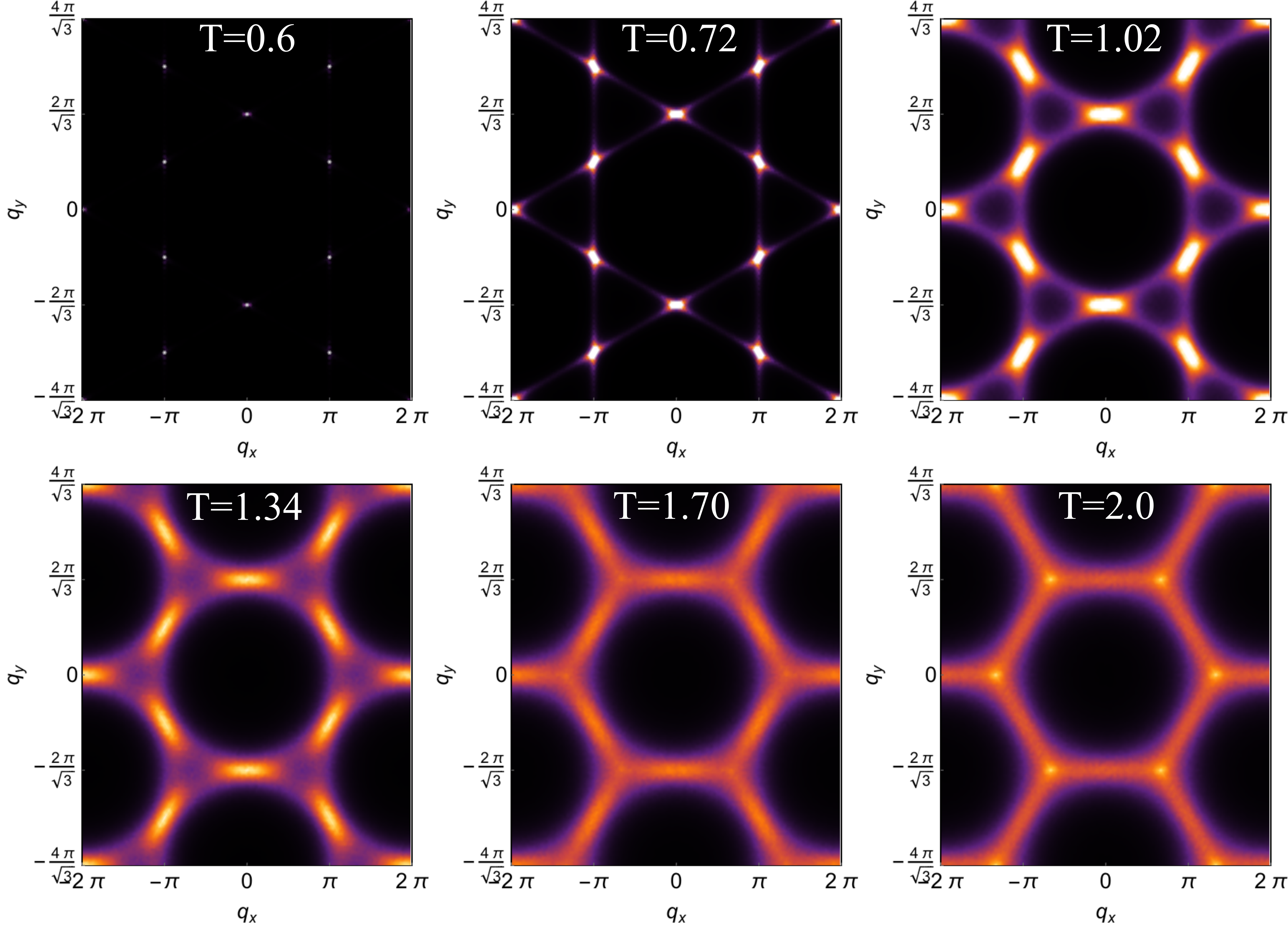}
\caption{\footnotesize{
The structure factor, $S({\bf q})$ [Eq.~2 in the main text] measured at $J_3/J_2=0.492$.
Low temperature Bragg peaks at ${\bf q} = (0,2\pi/\sqrt{3})$ and symmetry related wavevectors broaden above a phase transition ($T_1 \approx 0.7$) to the critical paramagnet.
This structure is typical of a system with large domains of stripe order.
As the domain size decreases the maxima broaden and eventually shift to ${\bf q} = (2\pi/3,2\pi/\sqrt{3})$ and symmetry related wavevectors.
}}
\label{fig:strucfacJ123}
\end{figure}

Monte Carlo simulations can also be used to measure the structure factor, $S({\bf q})$ [Eq.~2 in the main text], and an example is shown in Fig.~\ref{fig:strucfacJ123}.
At low temperature in the stripe state there are a set of Bragg peaks located at ${\bf q} = (0,2\pi/\sqrt{3})$ and symmetry related wavevectors.
Above the first order phase transition into the critical paramagnet these broaden and become diffuse.
This is characteristic of a system with large domain sizes.
As the domain size decreases the maxima broadens, and eventually shifts to ${\bf q} = (2\pi/3,2\pi/\sqrt{3})$ and symmetry related wavevectors.
The appearance of peaks in these positions is typical of the nearest-neighbour model, which is recovered at large $T$.

%%%%%%%%%%%%%%%%%%%%%%%%%%%%%%%%%%%%%%%%%%%%%%%%%%%%%%
% Tddw calculation
\section{The transition temperature T$_{\sf ddw}$}
\label{app:Tddw}
%%%%%%%%%%%%%%%%%%%%%%%%%%%%%%%%%%%%%%%%%%%%%%%%%%%%%%

The temperature, $T_{\sf ddw}$, at which the second-order phase transition between the stripe state and the nematic state occurs can be calculated exactly.
In Ref.~[\onlinecite{korshunov05}] this was done in the case of $J_2$ and $J_3$ interactions.
Here we show how this can be extended to also include $J_4$ and $J_5$ interactions.

Starting from the stripe state, a second-order transition to the nematic state occurs when the free energy of an isolated double domain wall (ddw) is equal to zero.
The energy cost of placing a ddw into the the stripe state has three contributions.
Assuming that the interactions between spins are zero beyond $J_5$, the energy cost per unit length of a single domain wall is given by $E_{\sf dw} = 2J_2-4J_3+4J_4$.
Furthermore parallel domain walls separated by a single lattice unit do not have an interaction energy.
Placing a corner in a ddw costs an energy $E_{\sf c} = 2J_2 -8J_4$ and two neighbouring corners have an interaction energy $E_{\sf c-c} = 4J_4$, while corners separated by more than a single lattice unit do not interact.
It can be seen that the $J_5$ interaction does not appear in any of these quantities, and thus will not affect $T_{\sf ddw}$ (see Fig.~5 in the main text).

The energy of a ddw of length $L$ can be mapped onto a 1-d Ising model, where the Ising variable $\tau_i^{\sf z} =1$ if there is a corner at site $i$ and $\tau_i^{\sf z} =-1$ otherwise.
Therefore the energy of a wall is given by,
\begin{align}
E_{\sf ddw} = 
2 E_{\sf dw} L + E_{\sf c} \sum_i \frac{\tau_i^{\sf z}+1}{2}
+ E_{\sf cc} \sum_i \frac{\tau_i^{\sf z}+1}{2} \frac{\tau_{i+1}^{\sf z}+1}{2}.
\end{align} 
The free energy can be determined using a transfer matrix, resulting in,
\begin{align}
\frac{\mathcal{F}_{\sf ddw}(T)}{L} = 
2 E_{\sf dw} 
- T \log\left[  \frac{1}{2} \left( 1+ e^{-\beta(E_{\sf c}+E_{\sf cc})} +\sqrt{1+4e^{-\beta E_{\sf c}} -2 e^{-\beta(E_{\sf c}+E_{\sf cc})} + e^{-2\beta(E_{\sf c}+E_{\sf cc})}  } \right) \right],
\label{eq:Tddw}
\end{align}
where $\beta = 1/T$ and the exact value of $T_{\sf ddw}$ results from solving $\mathcal{F}_{\sf ddw}(T_{\sf ddw})=0$.
In the case of only $J_2$ and $J_3$ interactions then $E_{\sf cc} = 0$, and Eq.~\ref{eq:Tddw} reduces to Eq.~5 of Ref.~[\onlinecite{korshunov05}].

%%%%%%%%%%%%%%%%%%%%%%%%%%%%%%%%%%%%%%%%%%%%%%%%%%%%%%
% J1-J2-J3-J4-J5 simulation details, snapshots and structure factor
\section{Triangular lattice Ising antiferromagnet with J$_2$, J$_3$ and J$_5$ interactions}
\label{app:J1J2J3J5}
%%%%%%%%%%%%%%%%%%%%%%%%%%%%%%%%%%%%%%%%%%%%%%%%%%%%%%

The triangular lattice Ising antiferromagnet with $J_2$, $J_3$ and $J_5$ interactions has been described extensively in the main text.
Here we show a set of representative snapshots taken from simulations peformed at $J_3/J_2=0.4$ and $J_5/J_2=0.5$.
\begin{figure}[h]
\centering
\includegraphics[width=0.99\textwidth]{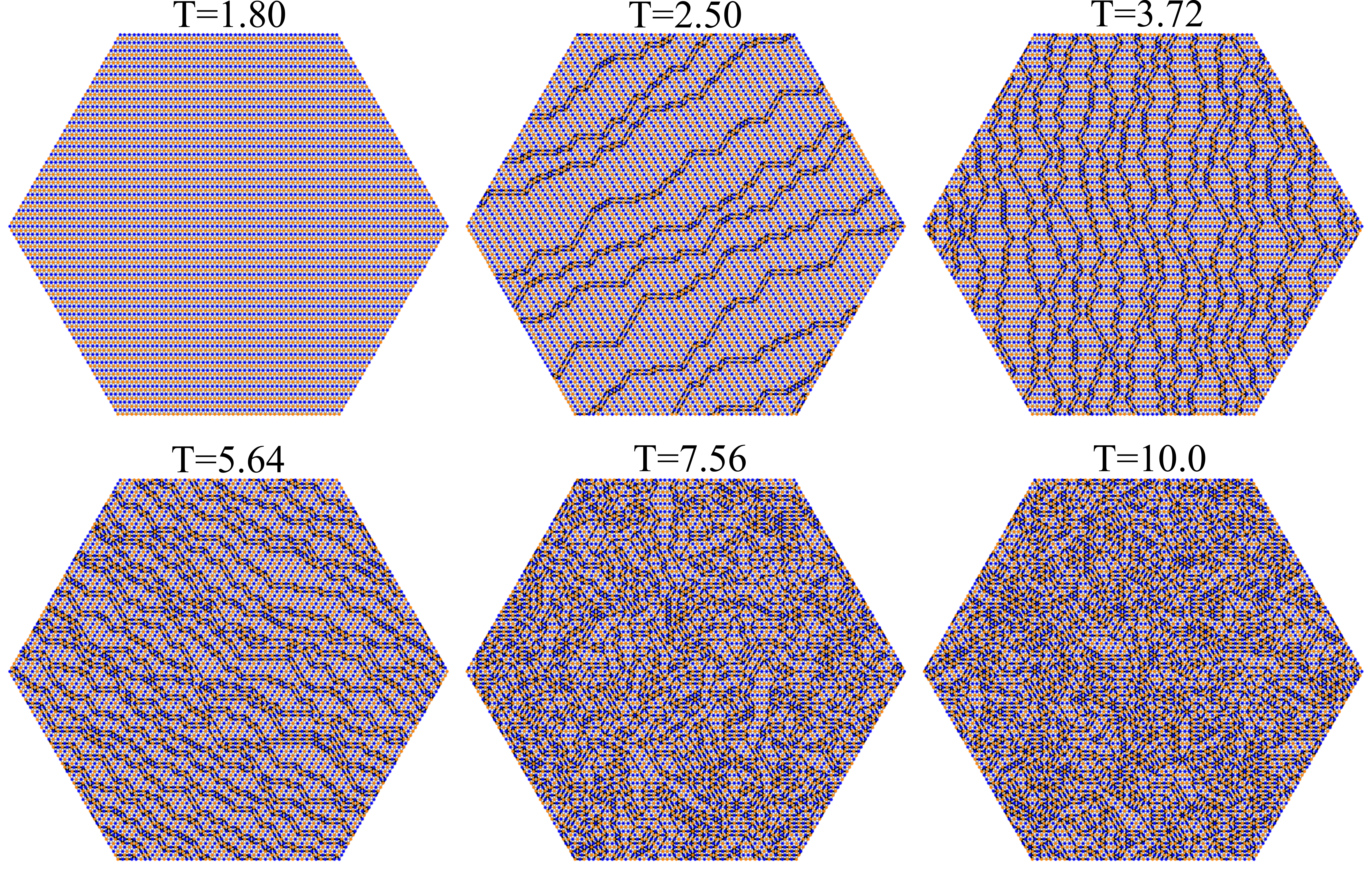}
\caption{\footnotesize{
Representative snapshots at a variety of temperatures for the triangular lattice Ising antiferromagnet with $J_3/J_2=0.4$ and $J_5/J_2=0.5$.
The system size is $L=48$.
Below $T_{\sf ddw}=2.29$ the system is in the stripe ground state, and all fluctuations are gapped.
Above this temperature, double domain walls condense into the system, and their density increases with increasing $T$.
Eventually at $T_{\sf 1}=5.67$ there is a first-order phase transition to the critical paramagnet, and even just above the transition the size of the stripe domains is small. 
}}
\label{fig:J1235snapshots}
\end{figure}

The snapshots, which are shown in Fig.~\ref{fig:J1235snapshots}, clearly show the difference between the three phases.
At $T<T_{\sf ddw} = 2.29$ the system is in the stripe state and no fluctuations are allowed.
At higher temperature double domain walls condense into the system, forming a nematic state, and it can be clearly seen that this breaks the $Z_3$ symmetry of the lattice.
Increasing the temperature within the nematic state increases the density of ddws.
Above the first-order phase transition into the critical paramagnet ($T_1=5.67$) the $Z_3$ symmetry is restored, and the domains of stripe order are small due to the relatively high temperature.

%%%%%%%%%%%%%%%%%%%%%%%%%%%%%%%%%%%%%%%%%%%%%%%%%%%%%%
% Beta=1/2 critical exponent
\section{Classical-quantum mapping and behaviour of nematic state close to T$_{\sf ddw}$}
\label{app:clas-quan}
%%%%%%%%%%%%%%%%%%%%%%%%%%%%%%%%%%%%%%%%%%%%%%%%%%%%%%

In order to study the behaviour of the nematic state in the vicinity of the second-order phase transition into the stripe state, we consider a classical-quantum mapping, following in spirit Refs.~[\onlinecite{pokrovsky79,pokrovsky80,villain81}].
The idea is to interpret the double domain walls (ddw) as fermions in 1 dimension.
The non-crossing constraint of the ddws is naturally built into the fermionic model due to the Pauli exclusion principle.
The second-order phase transition from the stripe to the nematic state is thus recast as a metal-insulator transition in a 1 dimensional model of fermions, and quantitative predictions can be extracted for the critical exponents.

\begin{figure}[h]
\centering
\includegraphics[width=0.75\textwidth]{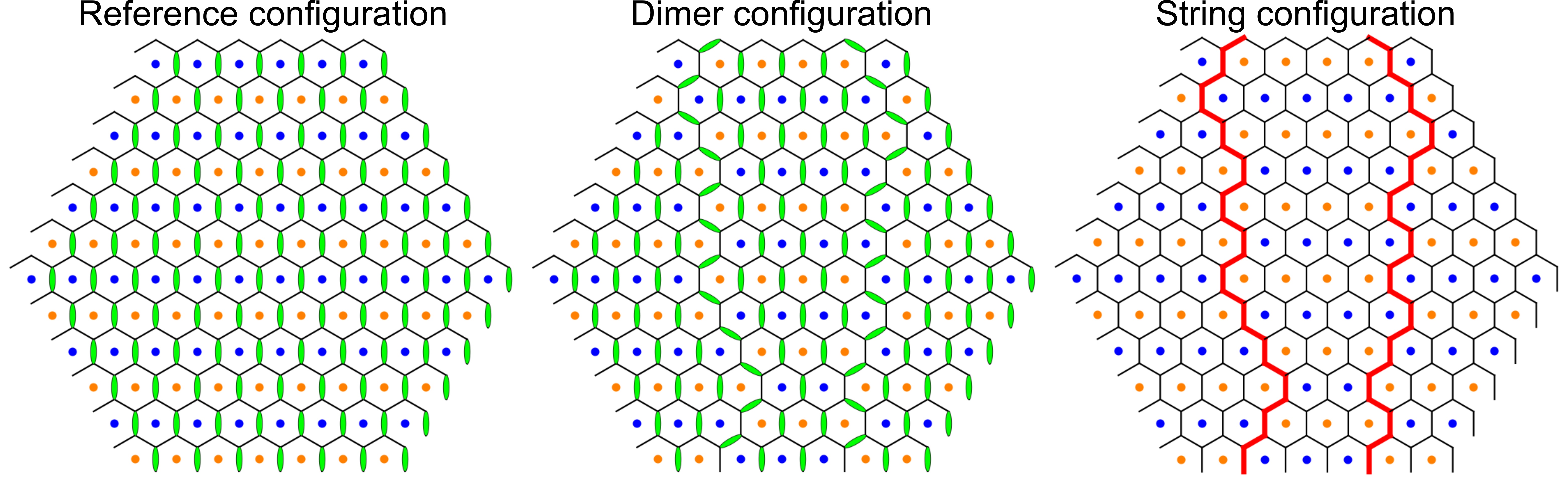}
\caption{\footnotesize{
Mapping from the ground states of the triangular lattice Ising antiferromagnet onto a string model on the honeycomb lattice.
The mapping from the Ising model to a dimer model on the honeycomb lattice is shown in Fig.~2 in the main text as well as Fig.~\ref{fig:dimerloops}.
A reference configuration of dimers is chosen as one of the stripe configurations.
A given dimer configuration is superimposed on this reference configuration, and honeycomb bonds on which these differ are assigned to belong to strings.
The strings (red lines) wind around the system, are directed parallel to the reference dimers and do not touch one another.
}}
\label{fig:stringconfig}
\end{figure}

There is a standard mapping between thermal phase transitions in $d$ dimensions and quantum phase transitions at zero temperature in $d-1$ dimensions.
The first step in performing such a mapping is to recast the triangular lattice Ising model in terms of a string model on the honeycomb lattice\cite{yokoi86,jiang06} [see Fig.~\ref{fig:stringconfig}].
Strings are defined by comparing a particular dimer configuration on the honeycomb lattice to a reference stripe configuration, which is taken to be one of the stripe states.
The difference between these gives the string configuration.
Strings form closed loops that wind the system, are directed parallel to the reference dimers and are forbidden from touching one another.
\begin{figure}[h]
\centering
\includegraphics[width=0.4\textwidth]{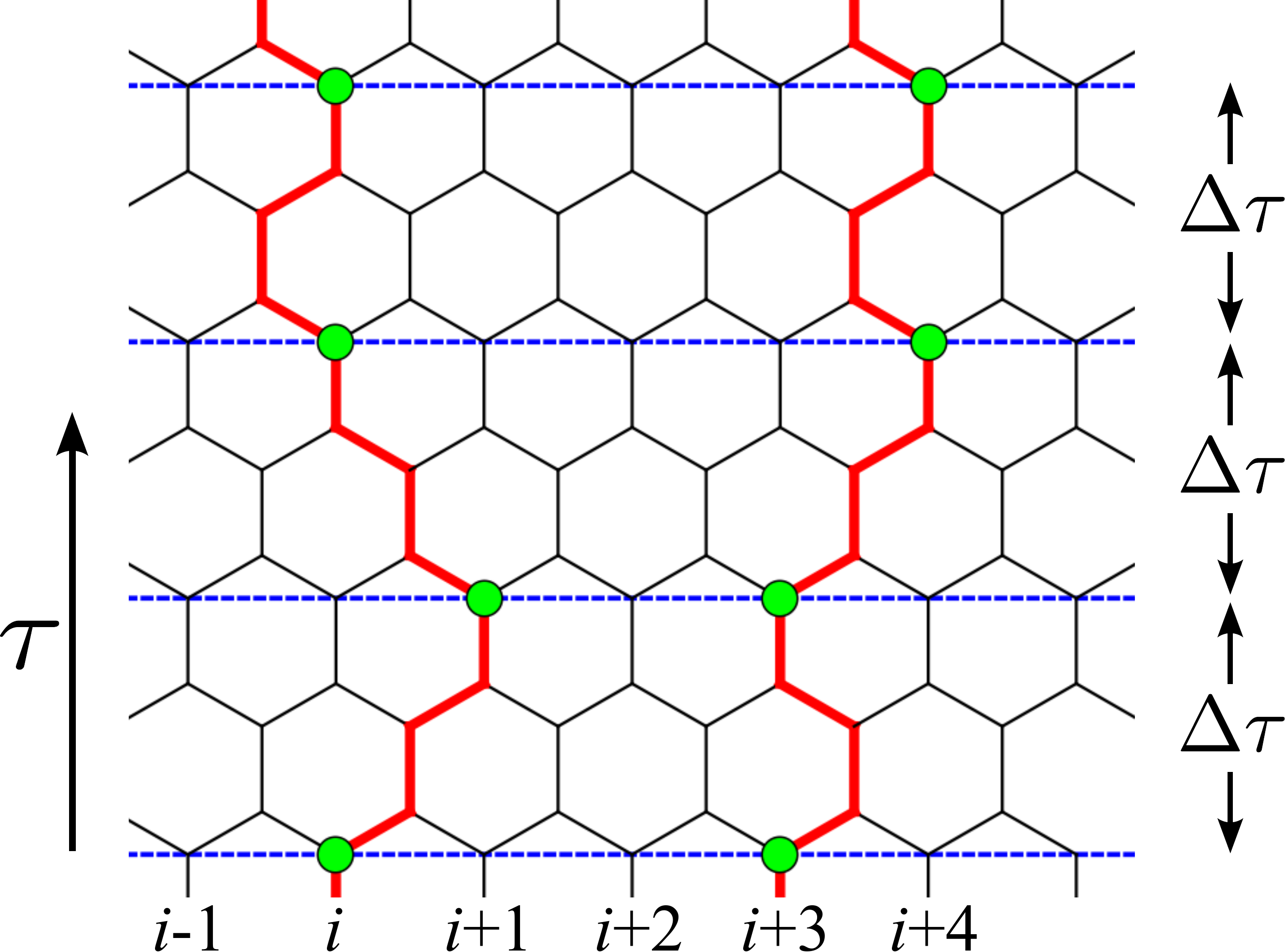}
\caption{\footnotesize{
Mapping from a 2 dimensional classical string model at finite temperature to a 1 dimensional zero-temperature quantum model of fermions.
Strings (red lines) are interpreted as the worldine of fermionic particles (green discs), with imaginary time elapsing in the direction parallel to the strings.
A single imaginary time step involves the string traversing four honeycomb bonds.
}}
\label{fig:fermionmap}
\end{figure}

Formally, a classical to quantum mapping is achieved by equating the transfer matrix, $\mathcal{T}$, with a quantum Hamiltonian, $\mathcal{H}$, according to $\mathcal{T} = e^{-\mathcal{H} \Delta \tau}$, where $\Delta \tau$ is a step in imaginary time.
Imaginary time is chosen to run parallel to the direction of the strings, and a string is considered to cross four honeycomb bonds in a single imaginary time step [see Fig.~\ref{fig:fermionmap}].
This construction is similar to the one carried out for spin ice in a [100] field\cite{jaubert08,powell08} and for a model of interacting dimers on the cubic lattice\cite{powell09}. 
In the quantum formulation the strings are interpreted as the worldlines of fermions.
This mapping respects the periodicity in imaginary time and the non-touching of neighbouring strings, which is built into the fermionic model due to the Pauli exclusion principle.

To determine an exact quantum Hamiltonian, $\mathcal{H}$, would be a very involved process, and we only consider the limit in which the string density is low.
In this dilute limit strings are equivalent to ddws, and therefore the approximation works well close to the second-order phase transition between the stripe and nematic states.
At leading order we ignore fermion-fermion interactions, as well as a history dependent asymmetry of the hopping amplitudes, and consider the 1-dimensional, free-fermion Hamiltonian,
\begin{align}
\mathcal{H} = 
\frac{1}{2} \sum_i  \left(c_i^\dagger  c_{i+1}^{\phantom\dagger} + c_{i+1}^\dagger  c_{i}^{\phantom\dagger} \right)
- (\mu-1) \sum_i  c_i^\dagger  c_{i}^{\phantom\dagger},
\end{align}
where the chemical potential, $\mu \propto (T-T_{\sf ddw})$.
Taking the Fourier transform results in,
\begin{align}
\mathcal{H} = 
\sum_k \xi_k c_k^\dagger  c_{k}^{\phantom\dagger} ,
\end{align}
where $\xi_k = \cos k+1 -\mu$.

\begin{figure}[h]
\centering
\includegraphics[width=0.5\textwidth]{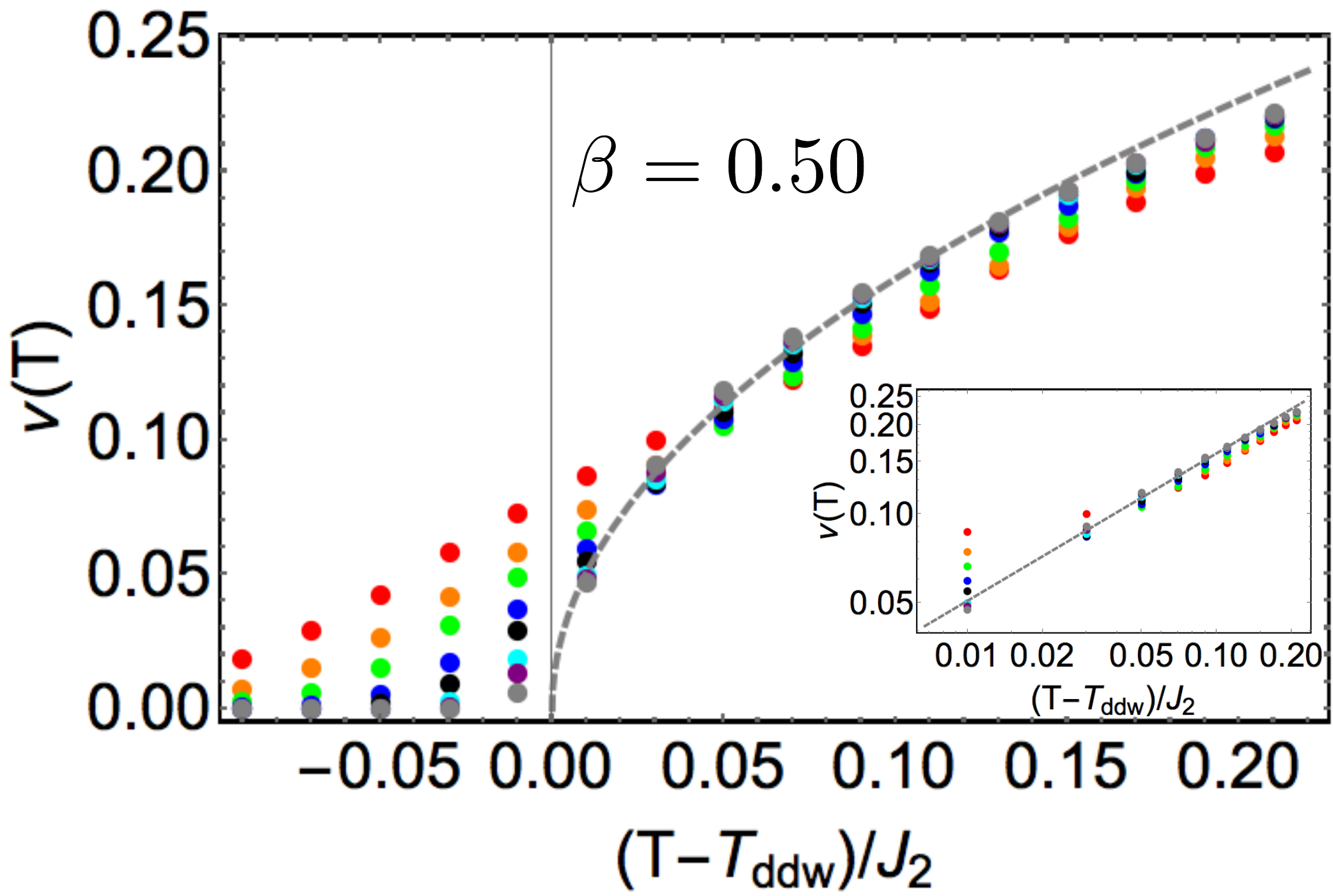}
\caption{\footnotesize{
Monte Carlo simulation results for the density of double domain walls (ddws), $\nu(T)$, close to the second-order phase transition between the stripe state and the nematic state (see Fig.~1 in the main text and Fig.~\ref{fig:J1235snapshots}).
Simulations are run with parameters $J_3/J_2=0.4$, $J_4=0$ and $J_5/J_2=0.5$ and for system sizes $L=24$ (red), $L=36$ (orange), $L=48$ (green), $L=72$ (blue), $L=96$ (black), $L=144$ (cyan), $L=192$ (purple) and $L=288$ (grey).
The dashed grey line shows the best fit to $\nu(T) \propto (T-T_{\sf ddw})^\beta$ for the $L=288$ data, which results in $\beta =0.50$, in agreement with the predictions of the free-fermion theory [see Eq.~\ref{eq:nu}].
The inset shows the same data and fit on a log-log scale.
Error bars are smaller than the point sizes.
}}
\label{fig:betahalf}
\end{figure}

This free fermion model exhibits a metal-insulator phase transition at $\mu=0$.
For $\mu < 0 $ the system is insulating since $\xi_k >0$ for all $k$ and the fermion band is empty.
For $\mu>0$ fermions start to partially fill the band close to $k=\pm \pi$ and the system is metallic.
The density of ddws in the classical system, $\nu(T)$ is equal to the density of fermions in the quantum system and simply calculating the filling ratio of the fermion band results in,
\begin{align}
\nu(T) \propto \frac{\pi}{2} - \arcsin (1-\mu).
\end{align}
Expanding this at small $\mu$ (small $T-T_{\sf ddw}$) gives,
\begin{align}
\nu(T) \propto (T-T_{\sf ddw})^\beta,
\label{eq:nu}
\end{align}
with $\beta = 1/2$.
This is in excellent agreement with Monte Carlo simulations performed close to $T_{\sf ddw}$ with system sizes up to $L=288$, as shown in Fig.~\ref{fig:betahalf}.
At higher temperatures the $\beta=1/2$ behaviour breaks down, since for a denser set of ddws it is no longer a good approximation to ignore fermion-fermion interactions.

The free fermion approximation can also be used to derive an expression for the structure factor, $S({\bf q})$ [Eq.~2 in the main text], in the nematic state.
This calculation is carried out in detail in Ref.~[\onlinecite{villain81}] and we do not reproduce it here (to apply to the case considered here one should set $q=0$ throughout Section~3 of Ref.~[\onlinecite{villain81}]).
One finds that there are a set of peaks at ${\bf q}_{\sf peak} = (\pm \pi \nu(T)/2, 2\pi/\sqrt{3} )$ and symmetry related wavevectors.
After Fourier transform, one finds that the correlation function has an algebraic decay, which is the signature of a critical phase.
In reciprocal space the free-fermion approximation leads to,
\begin{align}
S({\bf q}_{\sf peak} + \delta {\bf q)} \propto |\delta {\bf q}|^{\tau-2},
\label{eq:Sqcrit}
\end{align}
with $\tau=1/2$, and the constant of proportionality has a directional dependence.
Transforming this back to real space gives $S({\bf r}) \propto r^{-\tau}$ and for the direction perpendicular to the domain walls one can write, \mbox{$S({\bf r}) \propto \cos(\pi r/L) (L/r)^\tau$}.
\begin{figure}[h]
\centering
\includegraphics[width=0.5\textwidth]{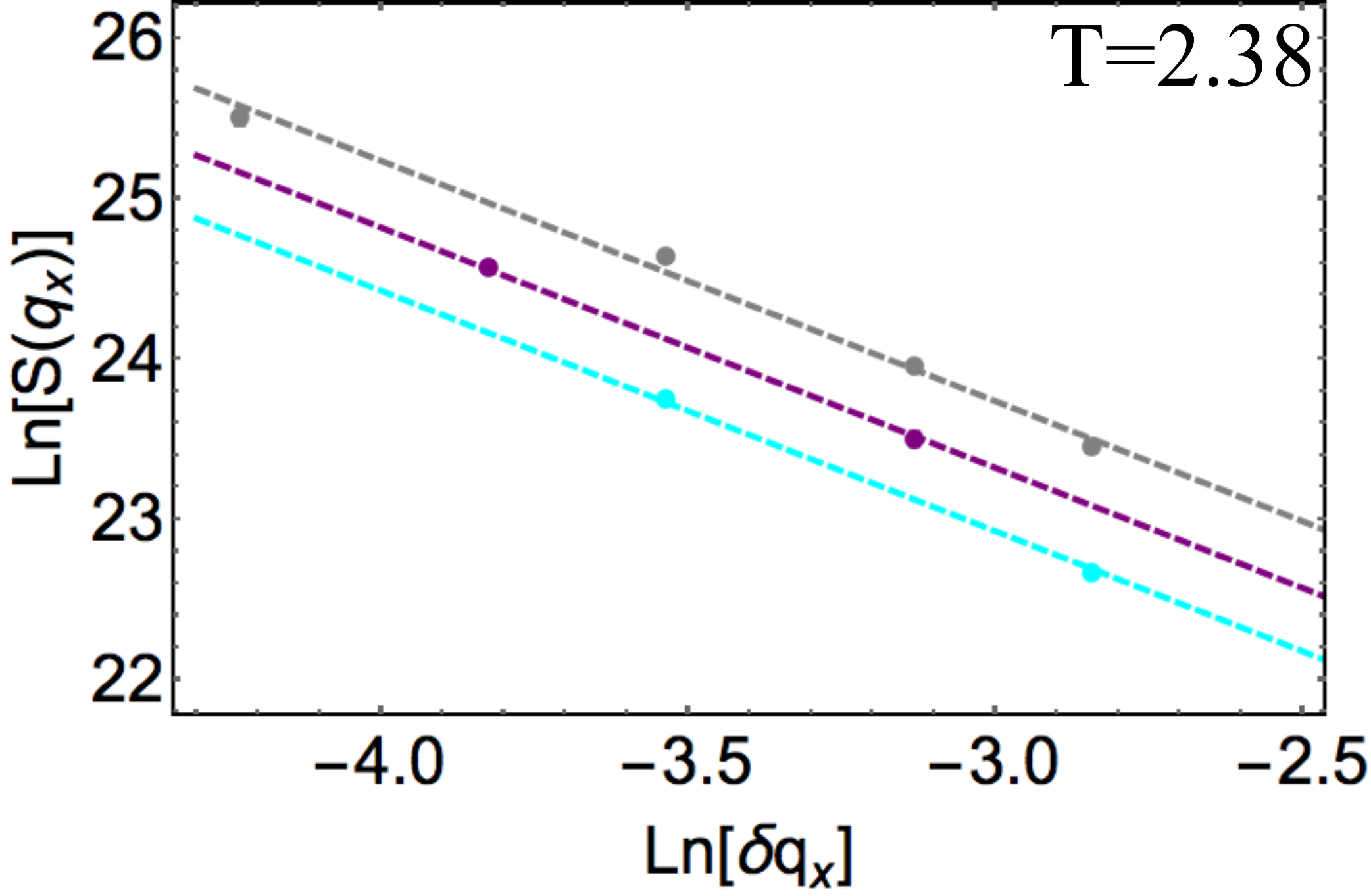}
\caption{\footnotesize{
Simulations of the structure factor, $S({\bf q})$ [Eq.~2 in the main text], with displacement $\delta q_{\sf x}$ from the peak at \mbox{${\bf q}_{\sf peak} = ( \pi \nu(T)/2, 2\pi/\sqrt{3} )$}.
This displacement is in the direction perpendicular to the domain walls. 
The parameters are $J_3/J_2=0.4$ and $J_5/J_2=0.5$ and $T=2.38$ ($T_{\sf ddw}=2.29$).
The system sizes are $L=144$ (cyan), $L=196$ (purple) and $L=288$ (grey).
The dashed lines show fits to the free-fermion prediction $\tau=1/2$ [see Eq.~\ref{eq:Sqcrit} and Ref.~[\onlinecite{villain81}]]. 
}}
\label{fig:tauSq}
\end{figure}

Monte Carlo simulations of the structure factor are shown in Fig.~3 of the main text.
While the nematic phase displays critical behaviour at all temperatures, the $\tau=1/2$ prediction is only relevant close to $T_{\sf ddw}$.
For higher temperatures we find $0<\tau<1/2$.
The region of reciprocal space over which the $\tau=1/2$ behaviour can be observed reduces as the temperature approaches $T_{\sf ddw}$, and it is therefore necessary to simulate large systems in order to have enough available ${\bf q}$ points.
For parameters $J_3/J_2 = 0.4$ and $J_5/J_2=0.5$ the critical temperature is $T_{\sf ddw} = 2.29$ and simulations at $T=2.38$ with system size $L=144$, $L=196$ and $L=288$ show evidence for $\tau=1/2$ critical behaviour, as shown in Fig.~\ref{fig:tauSq}.
%

%%%%%%%%%%%%%%%%%%%%%%%%%%%%%%%%%%%%%%%%%%%%%%%%%%%%%%
% edge states
\section{Edge excitations in the low-temperature stripe state}
\label{app:edges}
%%%%%%%%%%%%%%%%%%%%%%%%%%%%%%%%%%%%%%%%%%%%%%%%%%%%%%

The low-temperature stripe state of the triangular lattice Ising antiferromagnet is fluctuationless in the bulk.
This is because all excitations have a gap; local fluctuations involve the creation of pairs of defect triangles (triangles with all spins aligned), which cost $4J_1$ of energy, while non-local double domain wall (ddw) excitations have a free energy that is proportional to $L$.
In the limit $J_1\to \infty$ and the thermodynamic limit where $L\to \infty$ these types of excitations are forbidden in the stripe state.
However, fluctuations can occur at open boundaries.
These fluctuations are fractional and can freely move along the boundary, but are forbidden from penetrating the bulk.

\begin{figure}[h]
\centering
\includegraphics[width=0.8\textwidth]{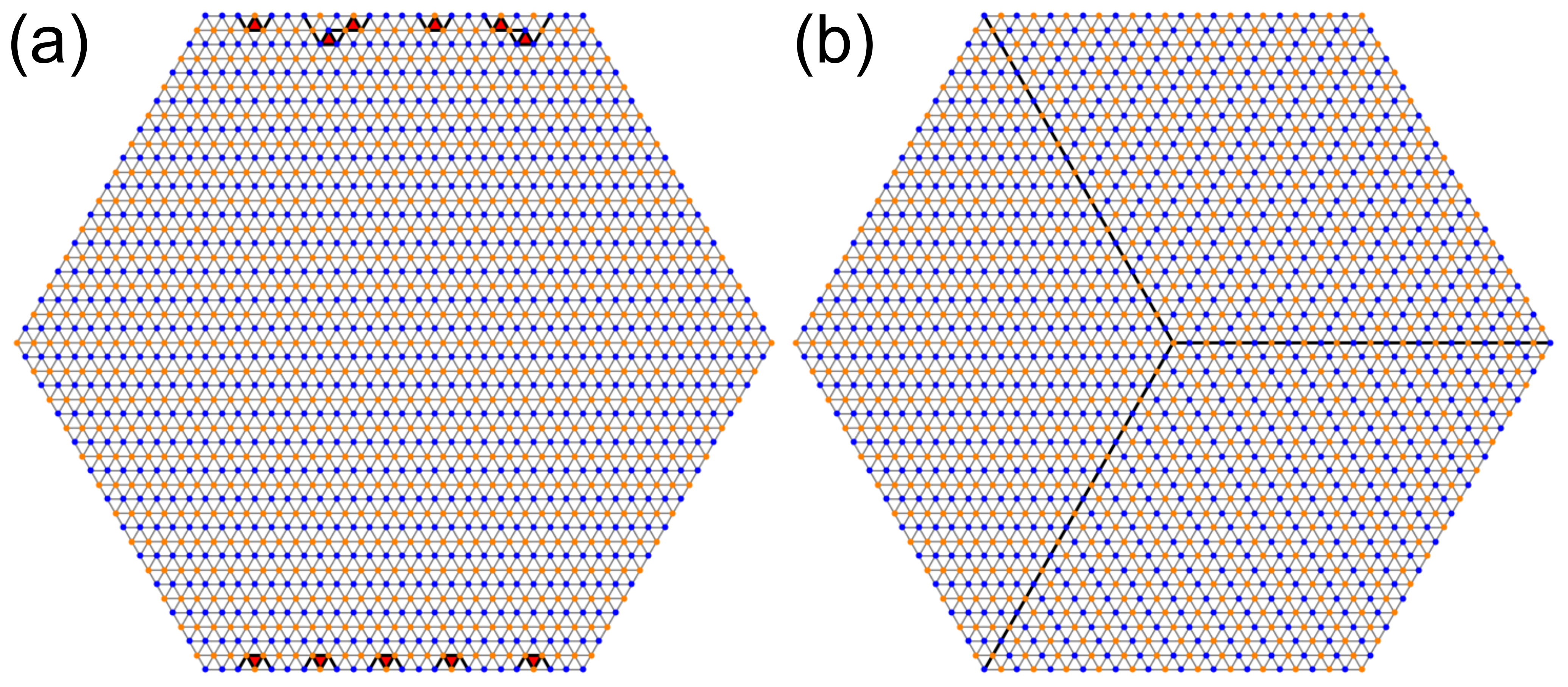}
\caption{\footnotesize{
Edge fluctuations in the low-temperature stripe state.
(a) A regular hexagon with open boundary conditions is prepared with stripes parallel to {\sf A} bonds, and Monte Carlo simulations are run with a local update scheme.
The parameters are $J_3/J_2=0.4$ and $J_5/J_2=0.25$.
At low temperature fractionalised edge excitations (red triangles) appear, and these only penetrate the bulk to a characteristic, temperature-dependent depth. 
(b) The ground state of the triangular lattice antiferromagnet with $J_1 \to \infty$.
In order to minimise the $J_1$ energy of the boundary, the system contains a minimal set of domain walls consistent with all boundaries having alternating spins. 
}}
\label{fig:edgefluc}
\end{figure}

For straight boundaries at $T=0$ there are two types of configurations, either stripes run parallel to the boundary and thus all the boundary spins are aligned, or stripes run at $60^\circ$ to the boundary and the boundary spins alternate.
Low-energy edge fluctuations occur only on boundaries with parallel stripes, since in this case each edge spin is surrounded by two up spins and two down spins, and can thus be reversed without a $J_1$ energy cost [see Fig.~\ref{fig:edgefluc}(a)].
The energy required to reverse an edge spin in such a situation is $E_{\sf edge} = 2J_2 -8J_3+4J_4$, and this creates a defect triangle.
This can be considered a fractional excitation, since in the bulk defect triangles always have to appear in pairs. 
Once created defect triangles can propagate freely along the boundary, but if they attempt to penetrate the bulk the energy cost grows proportionally to the penetration depth.

At $T=0$ different boundary configurations have different energies.
The relative energy cost per unit length between boundaries with alternating spins ($E^{\sf bound}_{60}$) and parallel spins ($E^{\sf bound}_{0}$) is given by \mbox{$E^{\sf bound}_0-E^{\sf bound}_{60} = J_1 - J_2 +3J_5$}.
For $J_1 \to \infty$ it is energetically favourable to have alternating spins on all boundaries, and the ground state will thus contain the minimal set of domain walls consistent with this requirement [see Fig.~\ref{fig:edgefluc}(b)].
However, if the system is prepared with a parallel-spin boundary, for example by cleaving a crystal, this state will be metastable as long as fluctuations are local.

In order to study this scenario we perform Monte Carlo simulations at low temperature on systems that are prepared with parallel-spin boundaries [see Fig.~\ref{fig:edgefluc}(a)].
We update the system by selecting spins that are surrounded at the nearest-neighbour level by equal numbers of up and down spins, and flipping them with a metropolis-type probability that depends on the $J_2 \dots J_5$ interactions.
Thus we fix the $J_1$ energy of the system.
As expected, we find that the number of defect triangles grows as the temperature is increased, and that the average penetration depth of the defect triangles into the bulk also increases.
%

%%%%%%%%%%%%%%%%%%%%%%%%%%%%%%%%%%%%%%%%%%%%%%%%%%%%%%
% edge states
\section{Fractional excitations in the stripe and nematic phases}
\label{app:fracexc}
%%%%%%%%%%%%%%%%%%%%%%%%%%%%%%%%%%%%%%%%%%%%%%%%%%%%%%

If the constraint $J_1/T \to \infty$ is relaxed, then the phase diagram of the triangular lattice antiferromagnetic Ising model (Fig.~1 in the main text) is qualitatively unchanged, as long as $J_1 \gg J_2$.
We have checked that this is the case for values of $J_1$ as low as $J_1=5J_2$.
The main difference is that for all $T>0$ there will be a finite, though possibly very small, density of defect triangles, and therefore the notion of a winding number is no longer well defined.

\begin{figure}[h]
\centering
\includegraphics[width=0.8\textwidth]{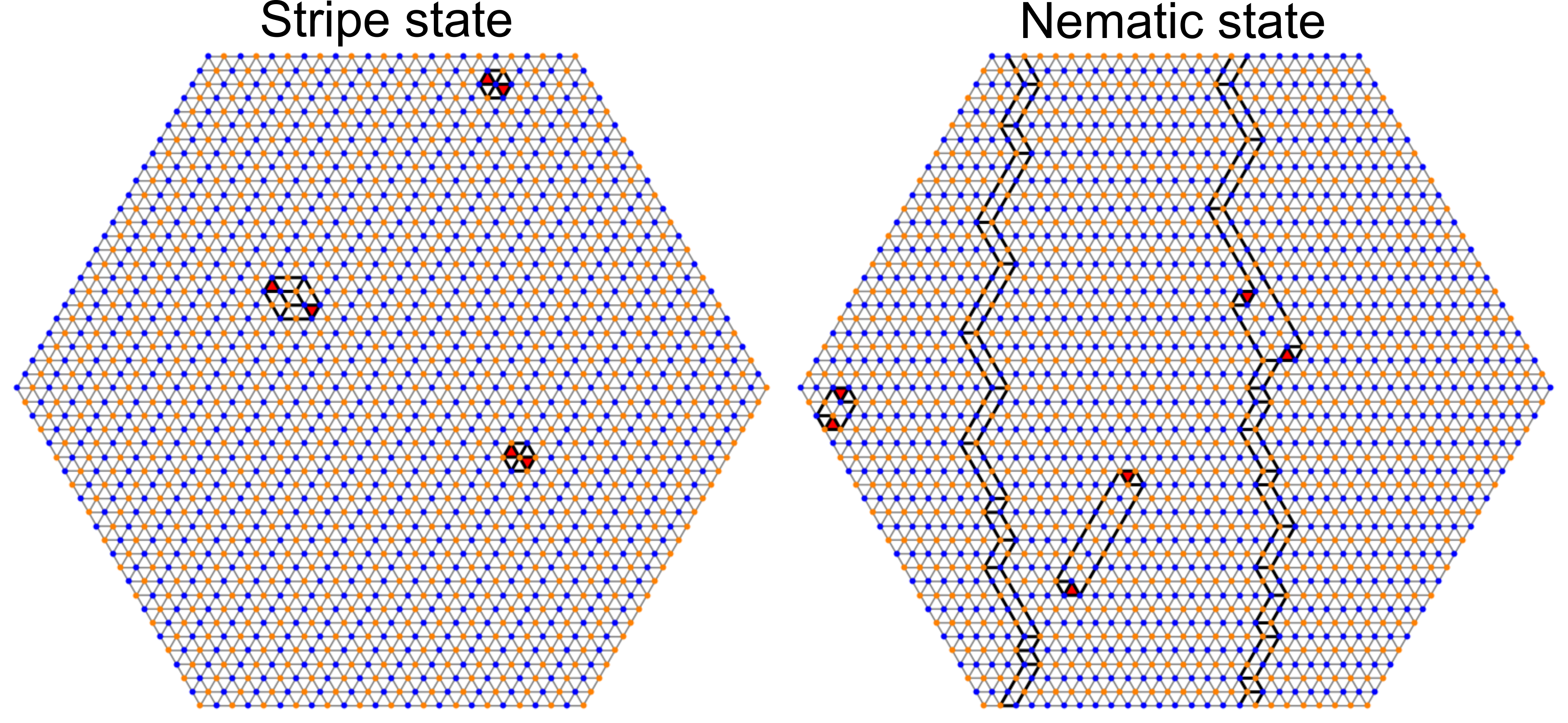}
\caption{\footnotesize{
Defect triangles in the stripe and nematic states (see Fig.~1 in the main text).
Since defect triangles (shown in red) are constrained to appear in pairs, individual triangles are fractional excitations.
In the stripe state (left) they are confined, since they are connected by a pair of double domain walls (or equivalently a double width double domain wall) and the free energy of such an object grow linearly with the separation.
In the nematic state (right) the free energy of a double domain wall goes to zero, and therefore defect triangles are deconfined.
The snapshots are taken from a simulation with $J_1/J_2 = 5$, $J_3/J_2=0.4$ and $J_5/J_2=0.5$.
}}
\label{fig:deftri}
\end{figure}

Defect triangles are always constrained to appear in pairs, and therefore individual defect triangles are fractional excitations.
In the stripe state they are confined, and a snapshot illustrating this is shown in Fig.~\ref{fig:deftri}.
It can be seen that pairs of defect triangles are linked by a pair of double domain walls (ddw), or equivalently a double width double domain wall.
In the stripe state this connection between the defect triangles has a free energy that grows linearly with the separation, confining the defects.

In the nematic state the defect triangles are deconfined, since the free energy of the pair of double domain walls connecting them goes to zero.
This is illustrated in Fig.~\ref{fig:deftri}.

%%%%%%%%%%%%%%%%%%%%%%%%%%%%%%%%%%%%%%%%%%%%%%%%%%%%%%
% height model
\section{Phenomenological analysis of the phase transition splitting}
\label{app:height}
%%%%%%%%%%%%%%%%%%%%%%%%%%%%%%%%%%%%%%%%%%%%%%%%%%%%%%

We demonstrate in the main text that the triangular lattice Ising antiferromagnet (TLIAF) with further neighbour interactions can either have a single first order transition or a double transition with an intermediate nematic state.
In order to gain a more universal understanding of when the transition splits, we here study a phenomenological model of the TLIAF.
This is based on mapping Ising configurations onto height configurations of the (111) face of a simple cubic crystal\cite{blote82}.

A microscopic height field, $h_i$, is defined on triangular lattice sites in the usual way\cite{blote82}.
Directional arrows are added to the bonds of the triangular lattice with an anticlockwise circulation around up-pointing triangles.
The origin is arbitrarily assigned the height $h = 0$, and then all other heights are fixed from considering the Ising configuration.
If two Ising spins are parallel, then the height changes by -2 in the positive bond direction, while if the spins are parallel the change is 1 in the positive bond direction.

The gradient of the height field on length scales large compared to the lattice spacing is related to the winding number sector.
In the paramagnet the $W=(0,0)$ sector dominates and the height field is rough, but has no gradient on a macroscopic scale.
In the stripe state the height field has the maximum possible gradient, orientated parallel to the stripe direction.

\begin{figure}[h]
\centering
\includegraphics[width=0.4\textwidth]{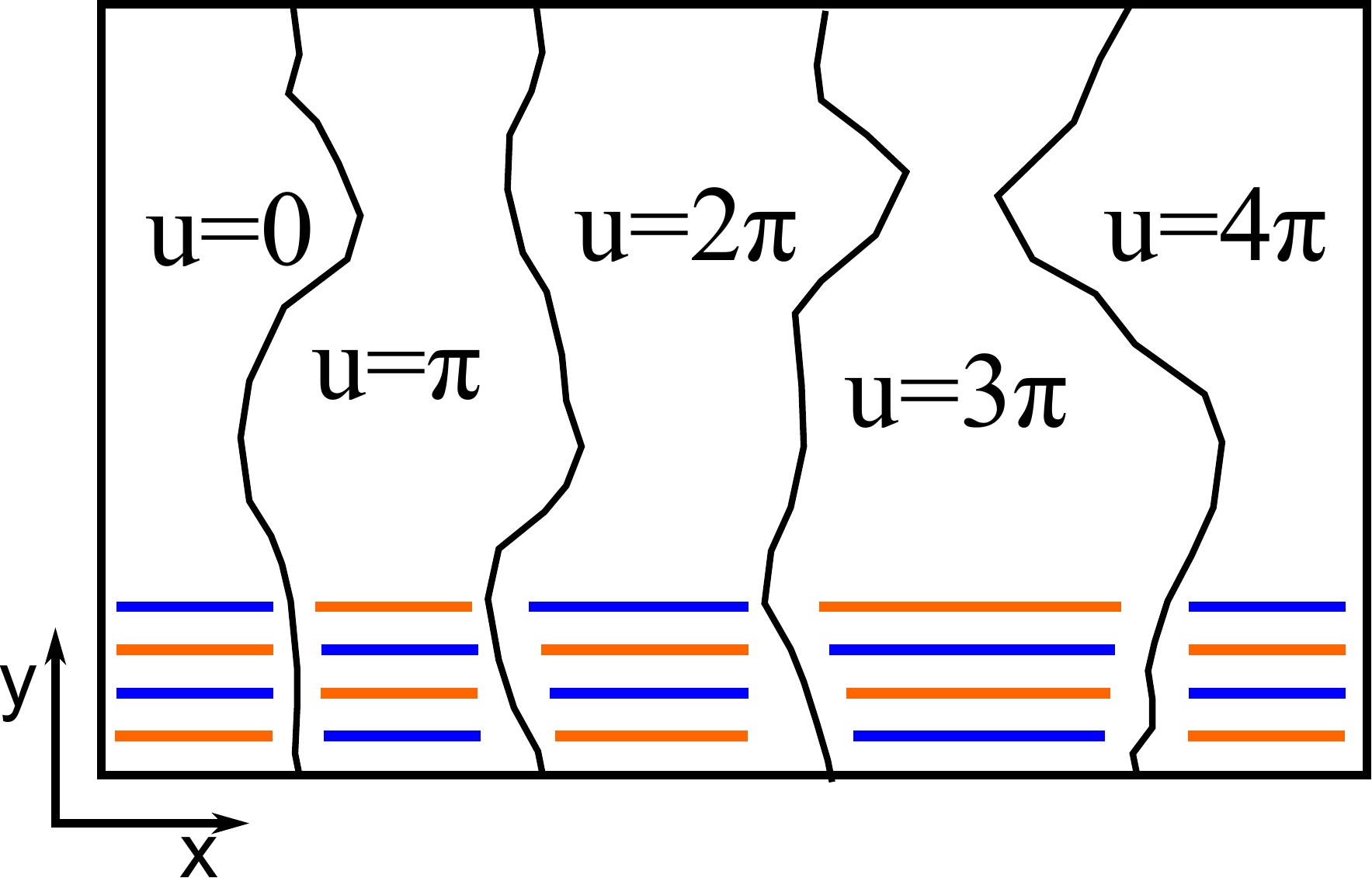}
\caption{\footnotesize{
The field $u({\bf r})$ [Eq.~\ref{fig:ufield}] in the nematic phase.
Double domain wall (black lines) separate stripe domains with opposite Ising spins (shown as blue and orange lines).
The field $u({\bf r})$ changes by $\pm \pi$ on crossing a double domain wall.
}}
\label{fig:ufield}
\end{figure}

It is useful to also define a field $u({\bf r})$ by,
\begin{align}
u({\bf r}) = \frac{\pi}{3} (h_{\sf str}({\bf r}) - h({\bf r})),
\label{eq:udef}
\end{align}
where $h_{\sf str}({\bf r})$ is the value of the height field in the stripe phase, with stripes parallel to the {\sf A} direction.
In the stripe phase $u({\bf r}) = 0$ and it changes by $\pm \pi$ on crossing a double domain wall (see Fig.~\ref{fig:ufield}),.

In order to study the long wavelength physics of $\mathcal{H}_{\sf Is}$ [Eq.~1 in the main text] above the upper phase transition, Ref.~[\onlinecite{korshunov05}] proposed the phenomenological free energy functional,
\begin{align}
\mathcal{F}_{\sf pm}[h] = \int d^2 r \left\{ 
\frac{K_2}{2} (\nabla h({\bf r}))^2  
+ K_3 \left[ ({\bf e}_{\sf A} \cdot \nabla) h({\bf r}) \right] 
\left[ ({\bf e}_{\sf B} \cdot \nabla) h({\bf r}) \right] 
\left[ ({\bf e}_{\sf C} \cdot \nabla) h({\bf r}) \right]  
+ \frac{K_4}{4} (\nabla h({\bf r}))^4
\right\},
\label{eq:Fheight}
\end{align}
where ${\bf e}_{\sf A} = (1,0)$, ${\bf e}_{\sf B} = (-1/2,-\sqrt{3}/2)$ and ${\bf e}_{\sf C} = (-1/2,\sqrt{3}/2)$ are the three positive bond directions.
Here $h({\bf r})$ arises from averaging the microscopic height fields over mesoscopic lengthscales, and making the continuum approximation.

At high temperature only nearest-neighbour interactions are important and therefore it is only necessary to consider the Gaussian term, which has stiffness $K_2 = \pi/9$.
The $W=(0,0)$ topological sector is thus favoured entropically.
As the temperature is lowered the value of $K_2$ reduces and eventually it becomes comparable with $K_3$ and $K_4$.
The parameter $K_3$ is related to the energy of a domain wall and it favours configurations with $|\nabla h| > 0$ pointing in one of the three lattice directions.
The $K_4$ term ensures that the free energy is bounded from below.

Analysis of $\mathcal{F}_{\sf pm}[h]$ [Eq.~\ref{eq:Fheight}] shows that there is a first-order phase transition at,
\begin{align}
K_{2{\sf c}_1} = \frac{K_3^2}{8K_4}.
\end{align} 
This transition breaks the $Z_3$ symmetry of the triangular lattice, and we assume that the symmetry is broken in such a way that the height gradient is a maximum parallel to the $x$ axis.
At the phase transition the gradient of the height field thus takes the value, 
\begin{align}
(\partial_{\sf x}h)_{{\sf c}_1}  = A_{{\sf c}_1} = -\frac{K_3}{2K_4}.
\end{align} 
Below the transition there are two possibilities, either a stripe state forms directly or there is an intermediate nematic state.
We make the assumption that it is the nematic state that forms, and then test the consistency of this assumption.

For $K_2<K_{2{\sf c}}$ we write the gradient of the height field as,
\begin{align}
\partial_{\sf x}h = A_{{\sf c}_1} +\delta (\partial_{\sf x}h).
\end{align} 
If we make the assumption that the two phase transitions are close in temperature the nematic state has a low density of double domain walls even close to the upper transition and therefore $\delta (\partial_{\sf x}h) \ll A_{{\sf c}_1}$ throughout the nematic state. 
Expansion of the free energy [Eq.~\ref{eq:Fheight}] in small $\delta (\partial_{\sf x}h)$ gives at second order,
\begin{align}
\mathcal{F}_{\sf nem}[h] = \int d^2 r \left\{ 
\frac{v_{\sf x}^\prime}{2} (\delta (\partial_{\sf x}h))^2  
+\frac{v_{\sf y}^\prime}{2}  (\partial_{\sf y}h)^2 
+\dots \right\}
\end{align}
where the zeroth and first order terms cancel and,
\begin{align}
v_{\sf x}^\prime &= K_2 + \frac{3}{2} K_3 A_{{\sf c}_1} +3K_4 A_{{\sf c}_1}^2 = K_2 \nonumber\\
v_{\sf y}^\prime &= K_2 - \frac{3}{2} K_3 A_{{\sf c}_1} +3K_4 A_{{\sf c}_1}^2 = K_2 + 12K_{2 {\sf c}_1}.
\end{align}
Thus the free energy cost of changing the height gradient becomes highly anisotropic between the two directions.
In terms of the field $u$ the free energy is,
\begin{align}
\mathcal{F}_{\sf nem}[u] = \int d^2 r \left\{ 
\frac{v_{\sf x}}{2} \left( \frac{\pi}{3}( A_{\sf str} -A_{{\sf c}_1})  - \partial_{\sf x}u \right)^2  
+\frac{v_{\sf y}}{2}  (\partial_{\sf y}u)^2 
+\dots \right\}
\end{align}
where $A_{\sf str} = \partial_{\sf x}h_{\sf str}$ and $v_i = 9v_i^\prime/\pi^2$.

We argue that in the nematic state it is not possible to ignore locking terms.
The reason for this is that in the absence of locking terms the field $u$ is energetically constrained to have a constant gradient, and thus double domain walls are not well defined.
Physically this is not consistent, since in the presence of a low density of domain walls, one would instead expect to have $u$ constant over large regions, separated by sudden jumps.
A locking term penalises regions in which $u$ deviates from a multiple of $\pi$, and therefore results in well defined domain walls.

The simplest way to add a locking term to the free energy is,
\begin{align}
\mathcal{F}_{\sf nem}[u] = \int d^2 r \left\{ 
\frac{v_{\sf x}}{2} \left( \frac{\pi}{3}( A_{\sf str} -A_{{\sf c}_1})  - \partial_{\sf x}u \right)^2  
+\frac{v_{\sf y}}{2}  (\partial_{\sf y}u)^2 
-V_0 \cos 2u \right\},
\end{align}
where the first term favours a nematic phase with average gradient $\partial_{\sf x}u = \frac{\pi}{3}( A_{\sf str} -A_{{\sf c}_1})$ and the last term favours the stripe state, since this has $\cos 2u =1$ for all $x$.

The free energy is now equivalent to that used to analyse the commensurate-incommensurate (C-IC) transition in systems where a gas is adsorbed onto a substrate (see for example the review~[\onlinecite{bak82}] and the many references contained within).
In adsorbed systems the locking term is the periodic potential of the substrate and the gradient term is typically written as $v_{\sf x}(\partial_{\sf x}u - \delta)^2/2$, where $\delta$ measures the mismatch between the substrate lattice constant and the preferred lattice constant of the adsorbed layer.
A C-IC transition takes place at \cite{schulz82},
\begin{align}
\delta =\frac{\pi}{3}(A_{\sf str} -A_{{\sf c}_1}) = \frac{4}{\pi} \sqrt{ \frac{V_0}{v_{\sf x}} }.
\end{align}
Therefore one expects a transition between the stripe and nematic states when,
\begin{align}
K_{2c_2} = \frac{144}{\pi^4} \frac{V_0}{( A_{\sf str} -A_{{\sf c}_1})^2}.
\label{eq:vxthreshold}
\end{align}
If $K_{2c_2} < K_{2c_1} $ then the above analysis is consistent with a double phase transition.
If $K_{2c_2} > K_{2c_1} $ there is a direct transition from the paramagnet to the stripe state.

%%%%%%%%%%%%%%%%%%%%%%%%%%%%%%%%%%%%%%%%%%%%%%%%%%%%%%

\end{document}